\documentclass[journal,twocolumn]{IEEEtran}
\usepackage{epsfig,color,amsmath,cite}
\usepackage{amsthm} 
\usepackage{amsmath}    
\usepackage[T1]{fontenc}
\usepackage[utf8]{inputenc}
\usepackage{bm}
\usepackage{epstopdf}
\usepackage{amssymb}
\usepackage{url}
\usepackage{enumitem} 
\usepackage{multirow}
\usepackage{hhline}
\usepackage{booktabs}
\usepackage{mathtools}
\usepackage{makecell}

\usepackage[linesnumbered,boxed,commentsnumbered,ruled,vlined,longend]{algorithm2e}
\makeatletter

\SetKwProg{Init}{Initialization}{}{Proceed}
\SetKwProg{St}{Start}{}{Proceed}
\makeatother
\usepackage{comment}

\DeclareMathOperator{\trace}{trace}

\DeclareMathOperator*{\argmax}{arg\ max}

\makeatother
\DeclareMathAlphabet\mathbfcal{OMS}{cmsy}{b}{n}

\newtheorem{mydef}{Definition}

\usepackage{stackengine}

\newcommand{\mat}[1]{\boldsymbol{#1}}
\renewcommand{\vec}[1]{\boldsymbol{\mathrm{#1}}}

\providecommand{\mA}{\ensuremath{\mat{A}}}
\providecommand{\mB}{\ensuremath{\mat{B}}}

\providecommand{\mF}{\ensuremath{\mat{F}}}

\providecommand{\mZ}{\ensuremath{\mat{Z}}}

\providecommand{\va}{\ensuremath{\vec{a}}}

\providecommand{\vu}{\ensuremath{\vec{u}}}

\providecommand{\vx}{\ensuremath{\vec{x}}}

\providecommand{\vz}{\ensuremath{\vec{z}}}

\newcommand{\m}{\boldsymbol}
\allowdisplaybreaks[4]
\usepackage[colorlinks = true,
linkcolor = blue,
urlcolor  = blue,
citecolor = blue,
anchorcolor = blue]{hyperref}

\newcommand{\mr}[1]{\mathrm{#1}}
\usepackage[framemethod=TikZ]{mdframed}
\mdfdefinestyle{MyFrame}{%
	linecolor=black,
	outerlinewidth=1.25pt,
	roundcorner=1.25pt,
	innerrightmargin=5pt,
	innerleftmargin=5pt,}
	
\usepackage{tikz}
\usetikzlibrary{matrix,positioning,decorations.pathreplacing}
\usetikzlibrary{shapes.geometric, arrows}
\usetikzlibrary{backgrounds}
\usetikzlibrary{shapes}
\usetikzlibrary{tikzmark}
\usetikzlibrary{calc}
\usetikzlibrary{arrows,shapes,positioning,shadows,trees,mindmap}
\usepackage[edges]{forest}
\usetikzlibrary{arrows.meta}
\colorlet{linecol}{black!75}
\usepackage{xkcdcolors} 

\usepackage[noabbrev]{cleveref}

\usepackage{mathtools}

\DeclarePairedDelimiter\abs{\lvert}{\rvert}%
\DeclarePairedDelimiter\norm{\lVert}{\rVert}%

\makeatletter
\let\oldabs\abs
\def\abs{\@ifstar{\oldabs}{\oldabs*}}
\let\oldnorm\norm
\def\norm{\@ifstar{\oldnorm}{\oldnorm*}}
\makeatother

\usepackage[english]{babel}
\usepackage[utf8]{inputenc}
\usepackage[super]{nth}

\usepackage{graphicx}
\usepackage{float}
\usepackage[caption = false]{subfig}

\usepackage{array}
\usepackage{threeparttable}

\usepackage[english]{babel}
\usepackage[utf8]{inputenc}
\usepackage[super]{nth}

\SetKwRepeat{Do}{do}{while}%

\everymath{\displaystyle}

\newcolumntype{E}{>{\centering\arraybackslash}m{0.5in}}
\newcolumntype{Q}{>{\centering\arraybackslash}m{3in}}
\newcolumntype{K}{>{\centering\arraybackslash}m{1.6in}}

\definecolor{airforceblue}{rgb}{0.36, 0.54, 0.66}
\usepackage[most]{tcolorbox}
\newtcbox{\colorboxouline}[1][]{boxsep=0pt,top=2.5pt,bottom=1pt,left=2pt,right=1pt,colframe=airforceblue,colback=airforceblue!10,boxrule=0pt,toprule=0pt,leftrule=3pt,sharp corners, enhanced jigsaw,,#1}

\usepackage[export]{adjustbox}
\usepackage{threeparttable}
\usepackage{multirow}
\usepackage{annotate-equations}
\usepackage{balance}

\usepackage{colortbl} 

\usepackage{mdframed}

\mdfdefinestyle{theoremstyle}{%
	linecolor=blue,linewidth=1pt,%
	 frametitlerule=true,%
	 frametitlebackgroundcolor=gray!20, innertopmargin=\topskip,}
	 
\usepackage[framemethod=TikZ]{mdframed}
\mdfsetup{skipabove=1mm,skipbelow=1mm}
\newcounter{theo}[section]

 	\definecolor{pansypurple}{rgb}{0.47, 0.09, 0.29}
 	\definecolor{darkcerulean}{rgb}{0.15, 0.38, 0.61}
 	\mdfdefinestyle{RepStyle}{%
 		skipabove=\topskip,
 		skipbelow=\topskip,
		innerleftmargin=0ex,
		innerrightmargin=0ex,
		innertopmargin=0ex,
 		linewidth=1.5pt,
 		leftline=false,
 		rightline=false,
 		frametitlefont={\color{darkcerulean}\normalfont\bfseries},
 		frametitlerule=false,
 		frametitlebackgroundcolor=white,
		backgroundcolor=white,
 		linecolor=darkcerulean,
 		theoremseparator={},
 		ntheorem=false 
 		}

\definecolor{darkblueilike}{rgb}{0.1843 0.4471 0.6196}

\definecolor{deepjunglegreen}{rgb}{0.0, 0.29, 0.29}
\newmdenv[  
topline=false,  
rightline=false,  
bottomline=false,  
leftline=true,  
linecolor=darkblueilike!90,  
linewidth=3pt,
innerleftmargin=1.5ex,  
backgroundcolor=darkblueilike!10,  
frametitle=CBSP,
frametitlefont={\color{darkblueilike!10}\normalfont\bfseries},
frametitlebackgroundcolor=darkblueilike!90
]{BStP}

\newmdenv[style=RepStyle]{Rep}

\definecolor{pinegreen}{rgb}{0.0, 0.47, 0.44}
\definecolor{upmaroon}{rgb}{0.48, 0.07, 0.07}
\definecolor{ruddybrown}{rgb}{0.73, 0.4, 0.16}
\definecolor{burntorange}{rgb}{0.8, 0.33, 0.0}

\definecolor{coolblack}{rgb}{0.0, 0.18, 0.39}
\definecolor{midnightgreen}{rgb}{0.0, 0.29, 0.33}
\definecolor{smokeytopaz}{rgb}{0.61, 0.77, 0.89}
\definecolor{copper}{rgb}{0.72, 0.45, 0.2}

\newcommand\mybox[2][]{\tikz[overlay]\node[fill=upmaroon!25,inner sep=2pt, anchor=text, rectangle, rounded corners=1mm,#1] {#2};\phantom{#2}}

\begin{document}

\title{\Large \textsc{Control Node Placement and Structural Controllability of Water Quality Dynamics in Drinking Networks}}

	\author{Salma M. Elsheri$\text{f}^{\dagger, \P, \ast \ast}$ and Ahmad F. Tah$\text{a}^{\dagger}$
		\vspace{-0.7cm}
		\thanks{$^\dagger$Department of Civil and Environmental Engineering, Vanderbilt University, Nashville, TN, USA. Emails: salma.m.elsherif@vanderbilt.edu, ahmad.taha@vanderbilt.edu.}
		\thanks{$\P$Secondary appointment: Department of Irrigation and Hydraulics Engineering, Faculty of Engineering, Cairo University.}
		\thanks{$^{\ast \ast}$Corresponding author.}
		\thanks{This work is supported by the National Science Foundation under Grants 2151392.}
		}
	 
	\maketitle

	\begin{abstract}
	Chlorine, the most widely used disinfectant, needs to be adequately distributed in water distribution networks (WDNs) to maintain consistent residual levels and ensure safe water. This is performed through control node injections at the treatment plant via booster stations distributed across the WDNs. While previous studies have applied various optimization-based approaches for booster station placement, many have failed to consider the coverage of the station injections and the dynamic nature of WDNs. In particular, variations in hydraulics and demand significantly impact the reachability and efficacy of chlorine injections which then impact optimal placement of booster stations. This study introduces a novel formulation that combines control- and graph-theoretic approaches to solve the booster station placement problem. Unlike traditional methods, our approach emphasizes maximizing the system's ability to control disinfectant levels with minimal control energy, taking into account the time-varying hydraulic profiles that lead to different optimal station placements. We propose a simple weighting technique to determine the placements by assessing the structural controllability of each configuration, based on the network's topology, independent of specific parameters like decay rates or pipe roughness. This method ensures effective chlorine coverage across the network. Our approach is validated on different networks, demonstrating its operational effectiveness, scalability, and practicality. 
	\end{abstract}
		

\markboth{Water Resources Research, In Press, December 2025}{} 

\section{Introduction and Paper Contributions}~\label{sec:Intro}
\IEEEPARstart{D}{isinfection} is a key process in the practice of maintaining water quality against bacterial growth and contamination spread in urban water systems. Operators of water distribution networks (WDNs) maintain sufficient disinfectant residuals, with chlorine being the focus of this study, across the network. This is achieved through \textit{(i)} controlled chlorine injections at the treatment plant and \textit{(ii)} the strategic placement of booster stations at a number of nodes within the network. The placement of booster stations is constrained by the accessibility of network nodes and the associated costs, limiting the number of stations that can be installed. 

The traditional approach to solving the chlorine booster station placement (CBSP) problem is to determine the geographic locations considering various objectives: minimizing the chlorine injection mass, maintaining residual levels, minimizing costs, reducing byproduct formation, and/or ensuring timely response to uncertain contamination events. However, WDNs are complex systems where real-time operations and varying consumer demands lead to changing flow rates and directions, directly influencing the reachability and capability of chlorine injections. Consequently, the CBSP problem yields different optimal locations for each hydraulic scenario. The conventional method for selecting final locations for fixed stations typically involves choosing the nodes that appear most frequently across various scenarios. To that end, the majority of studies in the CBSP literature overlook critical theoretical and practical considerations. They often fail to thoroughly assess the coverage and effectiveness of booster station injections, where coverage refers to the ability of booster stations to distribute chlorine effectively across the entire network. Additionally, these studies do not adequately examine how this coverage changes under varying hydraulic scenarios within the WDN. Furthermore, they lack a systematic approach for determining optimal station placements that ensures sufficient coverage across multiple scenarios while adapting to the network's dynamic nature.


In this paper, we introduce a novel control engineering-based CBSP formulation, aimed at maximizing system controllability. In this context, controllability is defined as the ability to effectively steer, regulate, and maintain disinfectant levels within the network to consistently meet the established water health standards. The placement strategy is designed to achieve this objective while minimizing chlorine injections, which in control terms corresponds to reducing the required control input energy, ensuring efficient and effective distribution across the network. In addition, we determine the final station placements by assessing the resulting system structural controllability---a concept that links system controllability to network topology, independent of specific underlying parameters (e.g., decay and reaction rates and pipe roughness coefficients). This is achieved through a graphical assessment of the network and its connectivity, thereby accounting for the unique layout of each network to maximize the coverage of chlorine injections over the network components. Moreover, we extend this approach to provide WDN operators with insights into backup locations for chlorine injection in cases of booster station malfunctions or insufficient capacity. The approach is applied and tested on various network sizes and scenarios, with considerations for scalability also discussed.

To the best of our knowledge, this is the first attempt to tackle the CBSP from a coupled control- and graph-theoretic perspective while also considering the operational aspects and practical considerations of WDNs. In the following sections, we review the existing literature and highlight the gaps that this study aims to fill.

\subsection{Literature Review}~\label{sec:LitRev}

The topic of CBSP in the water engineering field has a rich body of literature, which we cover in this section. To introduce our novel approach to solving this engineering problem, we survey the literature on the following: \textit{(i)} previous state-of-the-art methodologies to solve this problem and their limitations, \textit{(ii)} the concept of controllability and its application to place actuators (i.e., the controllers, which are the booster stations in our study), and \textit{(iii)} how this concept can be integrated with graph theory to perform a structural controllability analysis that considers the system's actual structure and identifies the most effective configuration of stations. 
This review provides the reader with both general and specific insights into how CBSP has been employed in the water engineering field, alongside the principles of general and structural controllability in control theory, and how these concepts can exploited and applied to this specific application for WDNs. Throughout this literature survey, we highlight the main gaps and drawbacks that this paper aims to address, leading to the presentation of its main contributions listed in Section \ref{sec:PaperCont}.

\vspace{0.1cm}
\noindent \textbf{Chlorine Booster Stations Placement.} 
The study \cite{boccelliOptimalSchedulingBooster1998a} has introduced the concept of placing booster stations along the network instead on relying on the station at the very start of the network. Since then, several studies have explored various methodologies to optimally allocate these booster stations, each contributing unique approaches but also presenting limitations. These methodologies differ in the formulation of the optimization problem and the techniques used to determine the locations of booster stations. The authors in \cite{sethEvaluationChlorineBooster2019} determine these locations by formulating and solving a mixed-integer linear program that minimizes the chlorine mass consumed by the population. 
The same approach is adopted by the authors in \cite{al-zahraniOptimizingDosageLocation2016} for optimal locations and scheduling of booster chlorination in a real water supply network located in Al-Khobar, Saudi Arabia. On the other hand, Propato \textit{et al.} \cite{propatoBoosterSystemDesign2004} have proposed a mixed-integer quadratic programming model to locate booster stations and determine their dosage schedules, optimizing the spatiotemporal distribution of chlorine residual across the network. 

The study \cite{pecciConvexHeuristicsOptimal2022} has investigated the problem of optimal placement and operation of valves and chlorine boosters by minimizing the average zone pressure while maintaining target chlorine concentrations. This problem is solved via a convex heuristic to generate candidate locations and evaluate configurations. Meng \textit{et al.} \cite{mengOptimizationBoosterDisinfection2011} optimize booster station locations from a hydraulic perspective using the particle backtracking algorithm, allowing placement at critical upstream nodes to meet disinfectant needs. Subramaniam \textit{et al.} \cite{subramaniamSetCoveringModels2000} have introduced the \textit{chlorine covering set theory}, which guides selecting optimal booster locations---ensuring chlorine needs are met for all nodes. 


Genetic algorithm (GA) have also been utilized to solve the CBSP problem 
by: \textit{(i)} minimizing the difference between chlorine concentration and the residual chlorine upper bound \cite{ostfeldConjunctiveOptimalScheduling2006a}, \textit{(ii)} taking into consideration the dispersion process in network's dead-ends \cite{abokifaOptimalPlacementOperation2018a}, \textit{(iii)} coupling the GA with a multi-species WQ model to optimize booster locations and dosage to minimize exposure to Escherichia coli (E. coli) \cite{islamMinimizingImpactsContaminant2017}, \textit{(iv)} linking the GA with EPANET-MSX to minimize the overall costs of booster stations placement, construction, and operation while delivering the water with acceptable residual chlorine and TTHM concentrations (i.e., disinfectant by-products) \cite{drewaOptimizedAllocationChlorination2007}, or \textit{(v)} integrating GA with particle swarm optimization techniques to optimize booster stations locations and scheduling to ensure regulatory compliance and minimize environmental impacts \cite{pinedasandovalOptimalPlacementOperation2021b}. Additionally, Behzadian \textit{et al.} \cite{behzadianNovelApproachWater2012a} have presented a two-phase multi-objective optimization approach for booster disinfection. The first phase determines booster locations by maximizing volumetric discharge with appropriate disinfectant levels and minimizing the total disinfectant mass. The second phase refines these locations to minimize discharge avoiding THM limits and maximize discharge with standard disinfectant levels. 


The main limitation in the studies reviewed in the preceding paragraphs is that the placement of booster stations is typically guided by objectives such as minimizing disinfectant use, maintaining residual concentrations, or reducing exposure to contaminants. 
While these approaches may assess the effectiveness of chlorine distribution, coverage---defined here as the extent to which chlorine injections can influence the network in a controllable and energy-efficient way---is often treated as a byproduct of the optimization results. However, it is rarely embedded directly into the optimization formulation with structural guarantees on system controllability or control energy. Additionally, the final placement locations are typically determined by considering either a single dominant hydraulic scenario or multiple scenarios. In the latter case, the final placement is often based on selecting the most picked locations across these different hydraulic scenarios. Our paper addresses the first limitation by formulating a placement problem that ensures submodularity, a control-theoretic concept discussed in subsequent sections, and effective distribution of the control energy. In other words, this CBSP problem is designed to choose locations that would result in maximizing the controllability with the application of minimal control input energy (i.e., chlorine injections). The second issue is resolved by evaluating numerous possible hydraulic scenarios and determining the placement through the application of a novel weighting technique across these scenarios. We employ a structural controllability-based weighting technique to ensure and assess not only the control energy required and the most apparent locations but also the maximum coverage of the injections. 
The following section summarizes the relevant literature on these topics and highlights their connection to the control and operation of drinking water networks.

\vspace{0.1cm}
\noindent\textbf{Controllability-Driven Actuators Placement.} In dynamic systems and control engineering, efforts have been put forth to tackle the actuator or control node placement problems by leveraging the concept of controllability \cite{summersOptimalSensorActuator2014}. System's controllability can be quantified using different metrics that reflect the size of the controllable subspace and the control energy stored \cite{pasqualettiControllabilityMetricsLimitations2014a}. That is, the actuator placement problem can be formulated by incorporating these metrics to obtain the optimal placements that result in maximizing the controllability over the system. Some of those metrics, satisfy an important property that is referred to as \textit{submodularity} \cite{summersSubmodularityControllabilityComplex2016b}. Submodularity reflects the principle of \textit{diminishing returns}. The {diminishing returns} property means that the benefit of adding an additional actuator decreases as more actuators are already placed. In other words, submodularity suggests that adding an actuator to a smaller subset of actuators typically results in a greater marginal increase in system controllability than adding the same actuator to a larger subset. The submodular nature of certain controllability metrics and formulating the actuator placement problem as set function optimization, allow for the use of greedy algorithms. These algorithms iteratively select the actuator that provides the most significant improvement in controllability, ensuring that the resulting configuration is within at least 63\% of the optimal placement. This bound comes from a well-established theoretical result, which states that for a monotone submodular function, a greedy selection process achieves at least $(1 - 1/e) \approx 63\%$ of the optimal value \cite{tzoumasMinimalActuatorPlacement2016,summersSubmodularityControllabilityComplex2016b}. In the context of WDNs, exploiting this concept is particularly beneficial for overcoming the complexity introduced by the discretization of WQ dynamics, which often leads to high-dimensional optimization problems where computing the absolute optimal solution becomes computationally infeasible under many scenarios and conditions.

The other layer of complexity that our paper tackles is considering different operational scenarios that are caused by the changing hydraulic settings in WDNs as flow rates and directions change due to varying consumer demands and other factors. These changes directly influence the system controllability and consequently the placement of booster stations \cite{elsherifWaterQualityControllability2024}. In our paper, we propose a weighting technique that is based on both graph and control theory to determine the final booster station configuration. In this technique, we assess the system's structural controllability, which determines whether the system’s graphical structure and connectivity allow for overall control of WQ dynamics through control input manipulation (i.e., chlorine injections), regardless of specific parameter values (e.g., flow velocities, pipe roughness coefficients, and decay and reaction coefficients) \cite{ching-tailinStructuralControllability1974,ramosOverviewStructuralSystems2022}. In other words, structural controllability focuses primarily on whether a node can be reached and influenced by control inputs, depending on how flow directions and hydraulic settings establish connectivity within the network. As a result, this technique favors booster station locations that maximize chlorine controllability and coverage for the given network topology, components, characteristics, and connectivity patterns resulting from different hydraulic scenarios. This graph-based approach recognizes that each WDN has its own distinct topology (e.g., branched, tree, looped, or composite networks), unique characteristics, and varying operational scenarios.

\subsection{Paper Contributions}~\label{sec:PaperCont}
The objective of this paper is to provide a control- and network-theoretic approach that determines the optimal geographic placements of chlorine booster stations. The detailed methodological (M--KC) and practical (P--KC) key contributions presented by our work are as follows. 

\noindent \mybox{\textbf{M--KC}}$\textcolor{upmaroon!25}{\blacktriangleright}$ 
This paper addresses the booster station placement problem based on a control- and graphical-theoretic approach. The CBSP problem is formulated to maximize the WQ controllability and minimize the energy required by the chlorine controlled inputs. This is achieved by incorporating controllability metrics and formulating the problem as a set function optimization—the problem variables are the sets of booster station locations. These metrics are selected for their important property, submodularity, which allows us to solve the set function optimization problem using a forward greedy algorithm that provides near-optimal placement with a guarantee of achieving a minimum percentage of the optimal solution.

\noindent \mybox[fill=burntorange!17]{\textbf{M\&P--KC}}$\textcolor{burntorange!17}{\blacktriangleright}$ 
We propose a dynamic and scenario-based booster station placement approach. We introduce a practical weighting technique that is based on system structural controllability and evaluates multiple common and realistic hydraulic scenarios and the resulting near-optimal placements. 

\noindent \mybox[fill=yellow!25]{\textbf{P--KC}}$\textcolor{yellow!25}{\blacktriangleright}$ The developed approach is tested on various network sizes, characteristics, topologies, and operational conditions. In addition, a scalability-driven framework is developed to expand this approach for large-scale networks and is tested on the C-town network. Furthermore, to bridge the gap between the findings of the theoretical algorithms and operational aspects, we provide actionable insights for operators, including utilizing our approach to determine backup locations for chlorine injections in scenarios where fixed stations fail or have insufficient capacity.

\begin{table*}[t!]
	\centering
	\caption{WQ models for different WDNs components\vspace{-0.3cm}}~\label{tab:WQModels}
	\vspace{-0.4cm}
	\begin{tabular}{l|c|c}
		\hline
		Component	&  WQ Model  & Eq. \\ \hline \hline
		Reservoir	& $
		\textcolor{upmaroon}{c_i^\mathrm{R}(t + \Delta t_\mathrm{WQ})} = \textcolor{blue}{1} \times \textcolor{upmaroon}{c_i^\mathrm{R}(t)} + \textcolor{pinegreen}{1} \times \textcolor{burntorange}{c_i^{\mathrm{B}_\mathrm{R}}(t + \Delta t_\mathrm{WQ})}$  &    \begin{subequations}~\label{eq:ReservWQ}\hspace{-0.1cm}\end{subequations}\eqref{eq:ReservWQ}  \\ 	\hline
		Tank 	&  $\begin{aligned}
			& \textcolor{blue}{V_i^\mathrm{TK}(t + \Delta t_\mathrm{WQ})} \textcolor{upmaroon}{c_i^\mathrm{TK}(t+ \Delta t_\mathrm{WQ})} = \textcolor{blue}{V_i^\mathrm{TK}(t)} \textcolor{upmaroon}{c_i^\mathrm{TK}(t)} + \sum_{j \in L_{\mathrm{in}}} \textcolor{blue}{q^j_\mathrm{in}(t)\Delta t_\mathrm{WQ}} \textcolor{upmaroon}{c^j_\mathrm{in}(t)}   \\ &  +  \textcolor{pinegreen}{V^\mathrm{B_\mathrm{TK}}_i(t+\Delta t_\mathrm{WQ})} \textcolor{burntorange}{c^\mathrm{B_\mathrm{TK}}_i(t+\Delta t_\mathrm{WQ})} -  \sum_{k \in L_{\mathrm{out}}} \textcolor{blue}{q^k_\mathrm{out}(t)\Delta t_\mathrm{WQ}} 	\textcolor{upmaroon}{c_i^\mathrm{TK}(t)}  + \textcolor{blue}{V_i^\mathrm{TK}(t) \Delta t_\mathrm{WQ}} R^\mathrm{TK}(\textcolor{upmaroon}{c_i^\mathrm{TK}(t)}) 
		\end{aligned}$	& \begin{subequations}~\label{eq:TankWQ}\hspace{-0.1cm}\end{subequations}\eqref{eq:TankWQ}   \\ 	\hline
		Junction 	& $	\textcolor{upmaroon}{c_i^\mathrm{J}(t)}= \frac{\sum_{j \in L_{\mathrm{in}}} \textcolor{blue}{q_{\mathrm{in}}^{j}(t)} \textcolor{upmaroon}{c_\mathrm{in}^j(t)}+\textcolor{pinegreen}{q^\mathrm{B_\mathrm{J}}_i(t)} \textcolor{burntorange}{c^\mathrm{B_\mathrm{J}}_i(t)}}{\textcolor{blue}{q^{\mathrm{D}_\mathrm{J}}_i(t)}+\sum_{k \in L_{\mathrm{out}}} \textcolor{blue}{q_{\mathrm{out}}^{k}(t)}}$	&  \begin{subequations}~\label{eq:JuncWQ}\hspace{-0.1cm}\end{subequations}\eqref{eq:JuncWQ}   \\ 	\hline
		Pump	& $\textcolor{upmaroon}{c_i^{\mathrm{M}}(t+\Delta t_\mathrm{WQ})} = \textcolor{blue}{1} \times \textcolor{upmaroon}{c_k^{\boldsymbol{\cdot}}(t+\Delta t_\mathrm{WQ})}$	&  \begin{subequations}~\label{eq:PumpWQ}\hspace{-0.1cm}\end{subequations}\eqref{eq:PumpWQ} 
		\\ 	\hline
		Valve	& $\textcolor{upmaroon}{c_i^{\mathrm{V}}(t+\Delta t_\mathrm{WQ})} = \textcolor{blue}{1} \times \textcolor{upmaroon}{c_k^{\boldsymbol{\cdot}}(t+\Delta t_\mathrm{WQ})}$ & \begin{subequations}~\label{eq:ValveWQ}\hspace{-0.1cm}\end{subequations}\eqref{eq:ValveWQ}  \\ 	\hline
		Pipe	& $		\begin{aligned}
			\textcolor{upmaroon}{\begin{bmatrix}
					c^\mathrm{P}_i(1,t+ \Delta t_{\mathrm{WQ}}) \\ c^\mathrm{P}_i(2,t+ \Delta t_{\mathrm{WQ}}) \\ \vdots \\ c^\mathrm{P}_i(l-1,t+ \Delta t_{\mathrm{WQ}}) \\ c^\mathrm{P}_i(l,t+ \Delta t_{\mathrm{WQ}}) 
			\end{bmatrix}} = \textcolor{blue}{(1-{\lambda}_i(t))} \textcolor{upmaroon}{\begin{bmatrix} c^\mathrm{P}_i(1,t) \\ c^\mathrm{P}_i(2,t) \\ \vdots \\ c^\mathrm{P}_i(l-1,t) \\ c^\mathrm{P}_i(l,t) \end{bmatrix}}  +\textcolor{blue}{{\lambda}_i(t)} 
			\textcolor{upmaroon}{\begin{bmatrix} c^\mathrm{J}_j(t) \\ c^\mathrm{P}_i(1,t) \\ \vdots \\ c^\mathrm{P}_i(l-2,t) \\ c^\mathrm{P}_i(l-1,t) \end{bmatrix}}
			+ \textcolor{blue}{\Delta t_{\mathrm{WQ}}} \begin{bmatrix} R^\mathrm{P}(\textcolor{upmaroon}{c^\mathrm{P}_i(1,t)}) \\ R^\mathrm{P}(\textcolor{upmaroon}{c^\mathrm{P}_i(2,t)}) \\ \vdots \\ R^\mathrm{P}(\textcolor{upmaroon}{c^\mathrm{P}_i(l-1,t)}) \\ R^\mathrm{P}(\textcolor{upmaroon}{c^\mathrm{P}_i(l,t)}) \end{bmatrix}
		\end{aligned}$	& \begin{subequations}~\label{eq:PipeWQ}\hspace{-0.1cm}\end{subequations}\eqref{eq:PipeWQ}  \\  \hline \hline
	\end{tabular}
\end{table*}

\noindent \textbf{Paper Organization.} The remainder of this paper is organized as follows: Section \ref{sec:WQModel} presents the WQ dynamics model and its state-space representation. Based on this representation, the notion of WQ controllability and its Gramian and metrics are introduced in Section \ref{sec:WQControl}. Following this, Section \ref{sec:BStPlace} formulates the booster station placement problem, which is then validated through several case studies encompassing various scales, layouts, and scenarios in Section \ref{sec:CaseStudies}. Finally, Section \ref{sec:ConcLimFW} provides conclusions, discusses the study's limitations, and recommends directions for future research.

\noindent \textbf{Notation.} Italicized, boldface upper and lower case characters represent matrices and column vectors: $a$ is a scalar, $\va$ is a vector, and $\mA$ is a matrix. The notation $\mathbb{R}^n$ denotes the sets of column vectors with $n$ real numbers, while $\mathbb{R}^{n \times m}$ denotes the sets of matrices with $n$ rows and $m$ columns. The variables with upper case characters $\boldsymbol{\cdot}^\mathrm{J}, \boldsymbol{\cdot}^\mathrm{R}, \boldsymbol{\cdot}^\mathrm{TK}, \boldsymbol{\cdot}^\mathrm{P}, \boldsymbol{\cdot}^\mathrm{M},$ and $\boldsymbol{\cdot}^\mathrm{V}$ represent the variables related to junctions, reservoirs, tanks, pipes, pumps, and valves.

\section{Water Quality Dynamics Model}~\label{sec:WQModel}
We model the WDN by a directed graph $\mathcal{G} = (\mathcal{N},\mathcal{L})$.  The set $\mathcal{N}$ defines the nodes and is partitioned as $\mathcal{N} = \mathcal{J} \cup \mathcal{T} \cup \mathcal{R}$ where sets $\mathcal{J}$, $\mathcal{T}$, and $\mathcal{R}$ are collections of junctions, tanks, and reservoirs. Let $\mathcal{L} \subseteq \mathcal{N} \times \mathcal{N}$ be the set of links, and define the partition $\mathcal{L} = \mathcal{P} \cup \mathcal{M} \cup \mathcal{V}$, where sets $\mathcal{P}$, $\mathcal{M}$, and $\mathcal{V}$ represent the collection of pipes, pumps, and valves. Following in this section, we introduce the governing equations of the WQ model to trace the chlorine concentrations at network components and its final state-space representation formulated over a simulation time period, referred to as $T_\mathrm{s}$, and at every WQ time-step $\Delta t_\mathrm{WQ}$. This WQ model is based on the principles of transport, mass balance, and single-species reaction dynamics. The single-species reaction dynamics are based on the assumption that chlorine decays linearly \cite{fisherEvaluationSuitableChlorine2011}.\footnote{We note the following about what we mean by \textit{linear}: the state-space, input-output dynamics form a linear dynamical model, not that the chlorine states themselves decay linearly. It is known that even a simple linear model (see for example~\eqref{eq:WQSS}) has states that evolve nonlinearly in time and space. Even with linear reaction models, the evolution of chlorine states is nonlinear yet the state-space model is linear. Furthermore, herein we do not consider the more accurate multi-species reaction and water quality dynamics as the focus of this paper on the methodological advancement of controllability. We thank Dr. Ian Fisher for pointing this out and seeking clarification.} 

These WQ model governing equations are summarized in Tab. \ref{tab:WQModels}. For reservoirs, concentrations are assumed to be unchangeable over time unless a booster station is located at the reservoir, with its injections mixed instantaneously as expressed in Eq. \eqref{eq:ReservWQ}---booster station's injections' concentration at Reservoir $i$ is denoted as $c_i^{\mathrm{B}_\mathrm{R}}$. Similarly, mixing in tanks and junctions is assumed to be instantaneous. That is, the concentration at Junction $i$ is calculated via Eq. \eqref{eq:JuncWQ}, with $q_{\mathrm{in}}^{j}(t)$ and $q_{\mathrm{out}}^{k}(t)$ representing the inflows and outflows from links connected to that junction, $q^{\mathrm{D}_\mathrm{J}}_i(t)$ as the consumers' demand,  $c_\mathrm{in}^j(t)$ as the inflow solute concentration, and $q^\mathrm{B}_i(t)$ as the flow of chlorine injected by booster station (if located) with concentration $c^{\mathrm{B}_\mathrm{J}}_i(t)$. Eq. \eqref{eq:TankWQ} drives chlorine concentration at Tank $i$ based on $V_i^\mathrm{TK}(t)$ being the water volume of the tank, and $V^\mathrm{B}_i(t+\Delta t_\mathrm{WQ})$ as the volume of the booster station's chlorine injection (if located) with $c^{\mathrm{B}_\mathrm{TK}}_i(t)$ concentration. Lastly, $R^\mathrm{TK}({c_i^\mathrm{TK}(t)}) $ is the single-species reaction expression in Tank $i$, which is equal to $R^\mathrm{TK}(c_i^\mathrm{TK}(t)) = - k_b c^\mathrm{TK}_i(t)$ with $k_b$ as the bulk reaction rate constant.

Pumps and valves in our model are considered as links with negligible length; accordingly, their concentrations are assumed to be equal to the the concentration at the upstream node---refer to Eq. \eqref{eq:PumpWQ} and Eq. \eqref{eq:ValveWQ}. Finally, the chlorine transport and reaction in pipes are modeled by the one-dimensional advection-reaction partial differential equation (PDE), which for Pipe $i$ is expressed as
\begin{equation}\label{equ:PDE}
	\partial_t c_i^\mathrm{P}=-{v_i(t)} \partial_x c_i^\mathrm{P} + R^\mathrm{P}(c_i^\mathrm{P}(x,t)),
\end{equation}
where $c^\mathrm{P}_i(x,t)$ is concentration in pipe at location $x$ along its length and time $t$; $v_i(t)$ is the mean flow velocity; and $R^\mathrm{P}(c^\mathrm{P}_i(x,t))$ is the single-species decay reaction expression. This reaction expression is formulated as $	R^\mathrm{P}(c^\mathrm{P}_i(l,t)) = -\Big(k_b+\frac{2k_{w}k_{f}}{r_{\mathrm{P}_i}(k_{w}+k_{f})}\Big) c^\mathrm{P}_i(l,t)$, where $k_{w}$ is the wall reaction rate constant; $k_{f}$ is the mass transfer coefficient between the bulk flow and the pipe wall; and $r_{\mathrm{P}_i}$ is the pipe radius.

In our model, Eq. \eqref{equ:PDE} is discretized over a fixed spatio-tamporal grid using the {Explicit Upwind scheme}, an Eulerian Finite-Difference based method \cite{korenRobustUpwindDiscretization1993,elsherifControltheoreticModelingMultispecies2023}. This scheme is conditionally stable, requiring the Courant-Friedrichs-Lewy condition to be satisfied. Specifically, the Courant number ${{\lambda}_i(t)}=\frac{{v_i(t)} \Delta t_\mathrm{WQ}}{\Delta x_i}$ must satisfy $0<{{\lambda}_i(t)} \leq 1$, for a given Pipe $i$. Consequently, Pipe $i$ with length $L_{i}$ is divided into segments such that the number of segments  $n_{{l}_i}=\Big\lfloor \frac{L_i}{{v_i(t)} \Delta t_\mathrm{WQ}} \Big\rfloor$ of length $\Delta x_i= \frac{L_i}{n_{{l}_i}}$. Note that, the symbol $\left \lfloor \cdot \right \rfloor $ denotes the floor function, which returns the greatest integer less than or equal to the input value. The chemical concentrations for the pipe segments, from the first segment $c^\mathrm{P}_i(1,t+ \Delta t_{\mathrm{WQ}})$ to all segments in between along the pipe's length, and reaching the last segment $c^\mathrm{P}_i(l,t+ \Delta t_{\mathrm{WQ}})$, are calculated as expressed in Equation \eqref{eq:PipeWQ}, assuming Junction $j$ is upstream of this pipe. 

The total number of states $n_x$ for the WQ models consists of the number of nodes $n_\mathrm{N}$ and links $n_\mathrm{L}$. The number of nodes includes reservoirs, junctions, and tanks, given by $n_\mathrm{N} = n_\mathrm{R} + n_\mathrm{J} + n_\mathrm{TK}$. Whilst, the number of links comprises pumps, valves, and the total of pipe segments, given by $n_\mathrm{L} = n_\mathrm{V} + n_\mathrm{M} + \sum_{i=1}^{n_\mathrm{P}} n_{{l}_i}$. The number of booster stations to be located at nodes, which are the system's controllers, is denoted as $n_u$.

To that end, the WQ single-species transport and reaction model can be formulated as a linear difference state representation, as expressed in \eqref{eq:WQSS}:

\begin{equation}~\label{eq:WQSS}
\hspace{-1.0cm}	\eqnmark[upmaroon]{vxtt}{\vx (t + \Delta t_\mathrm{WQ})} = \eqnmarkbox[blue]{mA}{\mA(t)} \eqnmark[upmaroon]{vxt}{\vx (t)} + \eqnmarkbox[pinegreen]{mB}{\mB(t)} \eqnmark[burntorange]{vu}{\vu(t)}. 
\end{equation}


In this representation, the state vector $\vx \in \mathbb{R}^{n_x}$ collects chlorine concentrations at all network components, while the control input vector $\vu \in \mathbb{R}^{n_u}$ gathers chlorine injection concentrations from booster stations. The state matrix $\mA \in \mathbb{R}^{n_x \times n_x}$ maps the time and space dependencies between the states according to the system's hydraulics, layout, and characteristics, and control input matrix $\mB \in \mathbb{R}^{n_x \times n_u}$ represents the influence of the control inputs on the state.
We have color-highlighted the variables and parameters in Tab. \ref{tab:WQModels} to match their corresponding placement in the color-coded representation in \eqref{eq:WQSS}.
\section{Water Quality Controllability Gramian and Submodular Metrics}~\label{sec:WQControl}
In this section, we discuss how to assess WQ controllability by formulating what is referred to as its Gramian and the associated metrics. These tools help quantify and evaluate controllability, which is utilized to formulate the CBSP problem. For the WQ model formulated as the linear state-space representation \eqref{eq:WQSS}, controllability is defined as the ability to steer and drive the chlorine concentration from specific initial values to a desired value by means of injections through chlorine booster stations (i.e., actuators) over a specific time period $T_p$ \cite{kalmanMathematicalDescriptionLinear1963,elsherifQualityAwareHydraulicControl2024}. To that end, the dynamic linear system \eqref{eq:WQSS} is said to be controllable if and only if the controllability matrix for $N_p= \frac{T_p}{\Delta t_{\mathrm{WQ}}}$ time-steps given as
\begin{multline}~\label{eq:control_matrix}
\small \hspace{-0.2cm}	\mathcal{C}_{N_{p}} := \{\begin{matrix}
		\mB, \;\; \mA \mB, \;\; \mA^{2} \mB, \dots, \;\; \mA^{N_p-1} \mB
	\end{matrix}\} \in \mathbb{R}^{n_{x}\times N_{p}n_{u}},
\end{multline}
is full row rank, i.e, $\mr{rank}(\mathcal{C}_{N_{p}}) = n_x$~\cite{hespanhaLinearSystemsTheory2018}, without loss of generality as we assume that  $N_{p}n_{u} > n_{x}$. This is known as Kalman rank condition~\cite{kalmanMathematicalDescriptionLinear1963}. 

However, matrix rank provides only qualitative binary metric---whether the system is controllable or not. Consequently, it often fails to quantitatively indicate the \textit{degree of system controllability} across various scenarios and cases. That being said, we consider more practical non-binary 
metrics that quantitatively measure WQ controllability in order to formulate and solve the CBSP problem. 
These metrics are based on the controllability Gramian $\m{W}_{c}(\m{A},\m{B},N_p):= \m{W}_{c} \in \mathbb{R}^{n_{x}}$ that is defined for $N_{p}$ sum of matrices pair $\m{A}$ and $\m{B}$ as
\begin{equation}~\label{equ:control_gram}
	\hspace{-0.5cm}	\m{W}_{c} := \sum_{\tau=0}^{N_p-1}\m{A}^{\tau}\m{B} \m{B}^{\top}(\m{A}^{\top})^{\tau} = 
	\mathcal{C}_{N_{p}}\mathcal{C}_{N_{p}}^{\top}.
\end{equation}
Matrix $\m{W}_{c}$ is non-singular if the system is controllable after time ${T_{p}}$, otherwise it is uncontrollable. 

Various quantitative controllability-energy-related metrics have been explored in the literature, each with different interpretations and properties \cite{pasqualettiControllabilityMetricsLimitations2014a,summersSubmodularityControllabilityComplex2016b}. This includes the $\trace(\m{W}_{c})$, $\trace^{-1}(\m{W}_{c})$, log determinant $\log \det(\m{W}_{c})$, and minimum eigenvalue $\lambda_{\min}(\m{W}_{c})$. Study \cite{elsherifWaterQualityControllability2024} assesses the
performance, practicality, and limitations of these metrics in the context of WQ dynamics.

\section{CBSP Problem Formulation}~\label{sec:BStPlace}

\vspace{-0.6cm}
\subsection{Set function, submodularity, and forward Greedy Algorithm}~\label{sec:CBSPDiffMetrics}

We formulate the CBSP problem to maximize WQ controllability by using the maximum number of stations allowed for the network, denoted by $n_s$. We refer to the set of booster station locations as the booster station set $\mathcal{S} \subset \mathcal{N}$, with $|\mathcal{S}| = n_s$. The objective function of this problem is expressed as a \textit{set function} $f$. For the given finite set of node $\mathcal{N}$, the set function $f : 2^\mathcal{N} \rightarrow \mathrm{R}$ assigns a real number to each subset of $\mathcal{N}$. In the context of our CBSP problem, $f$ represents the metric for how controllable the system is for a given set $\mathcal{S}$ of placements. 
{\begin{algorithm}[b!]
		\caption{Solving CBSP problem via Forward Greedy Algorithm}
		\label{alg:forward_greedy}
		\SetKwInOut{Input}{Input}
		\SetKwInOut{Output}{Output}
		\SetKwInOut{Init}{Initialize}
		\Input{Water network parameters and characteristics, hydraulic profile, number of booster stations $n_s$, and total simulation period $T_\mathrm{s}$}
		\Output{$\boldsymbol{\mathcal{S}} = \{\mathcal{S}(1), \ldots, \mathcal{S}(T_\mathrm{s})\}$ \textcolor{blue}{// Optimal booster station set for each hydraulic time-step} \
			$\mZ = [\vz(1), \ldots, \vz(T_\mathrm{s})]$ \textcolor{blue}{// Structural controllability check matrix}}
		\Init{$t = 1, \; \boldsymbol{\mathcal{S}} \gets \emptyset, \; \mZ=\varnothing$}
		\While {$t \leq T_\mathrm{s}$}{
			$\mathcal{S}(t) \gets \emptyset, \; \vz(t)=\varnothing$\\
			\While {$|\mathcal{S}(t)| \leq n_s$}{
				$\alpha \gets \argmax_{\alpha \in \mathcal{N}\backslash\mathcal{S}(t)}[f(\mathcal{S}(t)\cup \{\alpha\})-f(\mathcal{S}(t))]$ \newline \textcolor{blue}{// $f(\mathcal{S}(t))$ is calculated via \eqref{eq:BStPProbLogDet} or \eqref{eq:BStPProbTrace} } \\
				$\mathcal{S}(t) \gets \mathcal{S}(t) \cup \{\alpha\}$ \\
				$\beta = \text{\texttt{sc}}(\mA, \mB_\mathcal{S})$ \textcolor{blue}{// Binary variable: $1$ if structurally controllable, and $0$ if not} \\
				$\vz(t)=[\vz(t); \; \beta]$
			}
			$\boldsymbol{\mathcal{S}} \gets \boldsymbol{\mathcal{S}} \cup {\mathcal{S}}(t)$ \\
			$\mZ = [\mZ, \; \vz(t)]$ \\
			$t = t + \Delta t_\mathrm{H}$ 
		} 
		\Return $\boldsymbol{\mathcal{S}}$\;
\end{algorithm}}

In our study, we solve the problem twice, with each problem considering a different metric. These metrics are $\log \det (\m{W}_{\mathcal{S}})$ and $\trace (\m{W}_{\mathcal{S}})$, where $\m{W}_{\mathcal{S}}$ is the controllability Gramian associated with the booster station set $\mathcal{S}$. This $\m{W}_{\mathcal{S}}$ is calculated using \eqref{equ:control_gram}, with $\mB=\mB_\mathcal{S}$ built based on $\mathcal{S}$. We formulate two CBSP problems: one by using the $\log \det(\m{W}_{c})$ metric and the other by using the $\trace(\m{W}_{c})$. Therefore, the CBSP problems are expressed as follows

\vspace{0.1cm}
\begin{BStP}  
	\textcolor{darkblueilike}{$\log \det$-based:}
\begin{equation}~\label{eq:BStPProbLogDet}
		\begin{aligned}
		\underset{\mathcal{S} \subset \mathcal{N}}{\mbox{maximize}} \hspace{0.1cm} f(\mathcal{S})=\log \det (\m{W}_{\mathcal{S}}), \hspace{0.1cm}
			\mbox{subject to} \hspace{0.1cm} |\mathcal{S}| = n_s.
		\end{aligned}		
	\end{equation}
	
\noindent \textcolor{darkblueilike}{$\trace$-based:}
		\begin{equation}~\label{eq:BStPProbTrace}
		\begin{aligned}
			\underset{\mathcal{S} \subset \mathcal{N}}{\mbox{maximize}} \hspace{0.1cm} f(\mathcal{S})=\trace (\m{W}_{\mathcal{S}}), \hspace{0.1cm}
			\mbox{subject to} \hspace{0.1cm} |\mathcal{S}| = n_s.
		\end{aligned}		
	\end{equation}
\end{BStP}

\begin{figure*}[t!]
	\centering
	\includegraphics[width=0.9\textwidth]{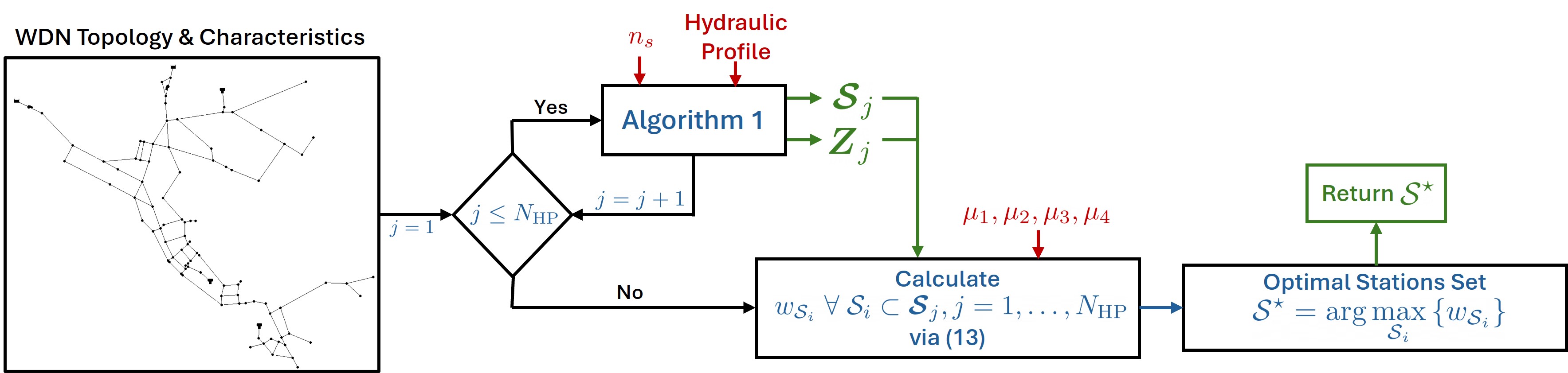}
	\caption{Flowchart of the chlorine booster station placement problem solution.}~\label{fig:FC} \vspace{-0.5cm}
\end{figure*}

While both metrics relate to control energy (energy in chlorine injections), the $\trace(\cdot)$ and $\log \det (\cdot)$ offer distinct insights into system controllability. The $\trace$ metric is inversely related to the average energy in all directions. Whilst, the $\log \det$ metric provides a volumetric measure of the reachable state space with one unit or less of the input energy. That being said, each of these metrics drives the CBSP problem towards achieving higher system controllability while minimizing the control energy but with different emphases and distributions of these aspects across the system, often leading to different solutions in various scenarios.

Solving this problem, specifically for large-scale networks is extremely computationally demanding. This issue is addressed by exploiting the important property of the objective set functions in \eqref{eq:BStPProbLogDet} and \eqref{eq:BStPProbTrace}: \textit{submodularity} \cite{summersSubmodularityControllabilityComplex2016b,summersOptimalSensorActuator2014}. We refer the readers to reference \cite{lovaszSubmodularFunctionsConvexity1983} for the mathematical definition of submodular functions. In abstract definition, submodularity is a diminishing returns property where adding an element to a smaller set gives a larger gain than adding one to a larger set. Thereby, this property and its structure, allows us to employ forward greedy algorithm \cite{krauseSubmodularFunctionMaximization2014,nemhauserAnalysisApproximationsMaximizing1978} to obtain near-optimal placements while being computationally tractable. We solve these CBSP problems at every hydraulic time-step, $T_p=\Delta t_\mathrm{H}$, and obtain the optimal placements, which allows the formulation of the controllability matrix \eqref{eq:control_matrix} and Gramian \eqref{equ:control_gram} to have time-invariant representations, as $\mA$ and $\mB$ remain time-invariant within this hydraulic time-step. Algorithm \ref{alg:forward_greedy} lists the details of applying the forward greedy algorithm and obtain the optimal booster station locations for the whole simulation period. 

\begin{figure*}[t!]
	\centering
	\includegraphics[width=0.88\textwidth]{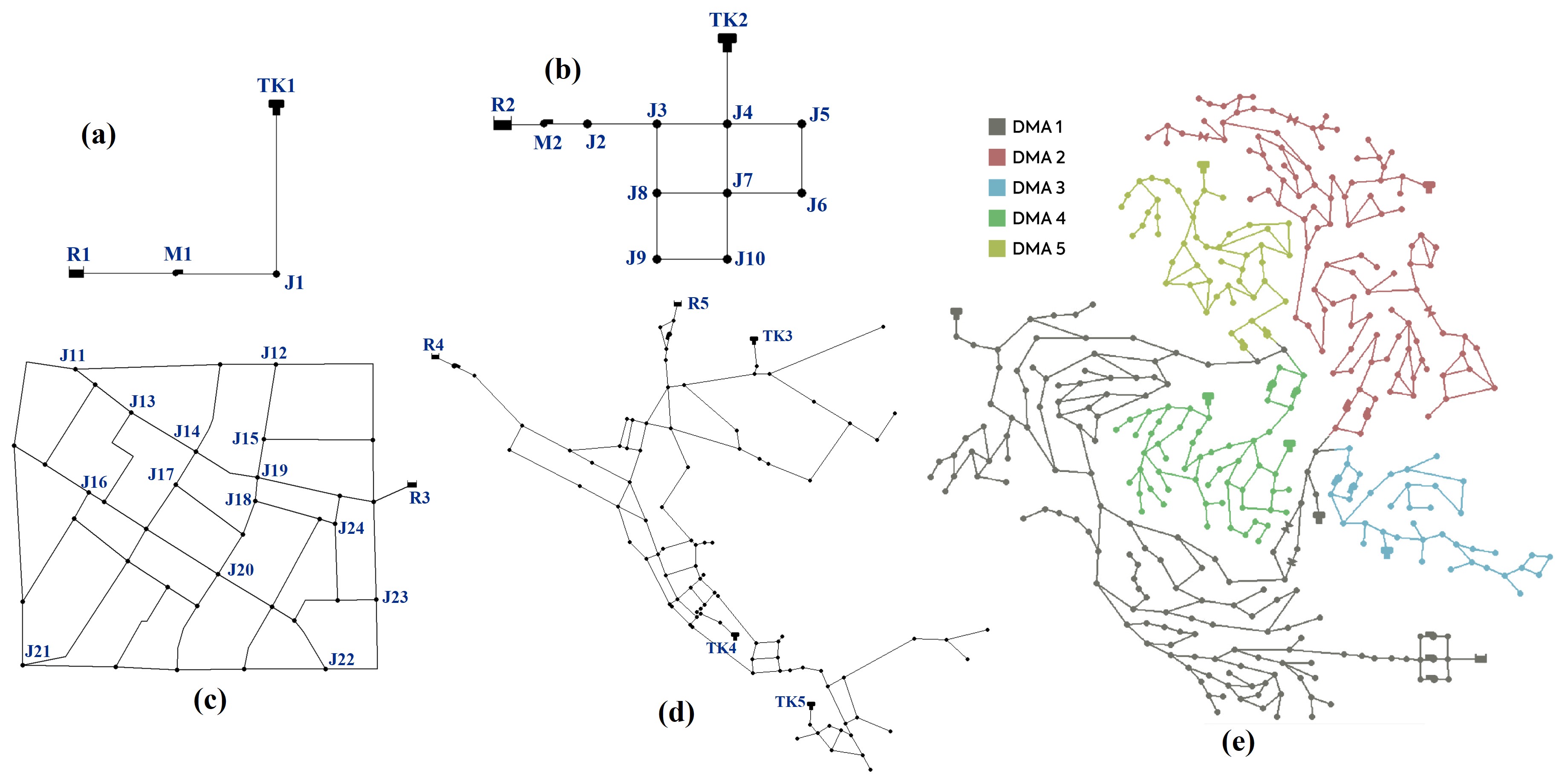}
	\vspace{-0.4cm}
	\caption{Test networks: (a) Three-node network, (b) Net1, (c) FOS, (d) Net3, and (e) C-Town network.}~\label{fig:CaseStudy} \vspace{-0.5cm}
\end{figure*}

While we formulate this problem and solve it in Algorithm \ref{alg:forward_greedy} with the entire set of nodes $\mathcal{N}$ are candidates for the CBSP placement, in many WDNs, some nodes may be inaccessible or have restrictions for chlorine injections. This approach is flexible enough to exclude those nodes from the selection pool and consider only actual candidates by changing the set $\mathcal{N} \backslash \mathcal{S}(t)$ in Step 4 to $\mathcal{W} \backslash \mathcal{S}(t)$, where $\mathcal{W}$ is the updated candidate nodes.

\subsection{Strategies for Booster Station Allocation Considering Different Hydraulic Profiles}~\label{sec:BStVsHydraulic}
In this section, we introduce strategies for obtaining the final booster station allocation by developing a weighting technique primarily based on the system's structural controllability. We consider structural controllability because the loss of controllability is often due to structural reasons. Essentially, the connections between inputs, states, and outputs are not strong enough to ensure the system's controllability. Through structural analysis, we can assess how much of the system's controllability depends solely on the presence of these connections between inputs and outputs. In other words, we evaluate these properties without relying on specific parameter values such as pipe coefficients, reaction rates, etc.

To that end, based on the dynamical system in \eqref{eq:WQSS} we define the class $\mathcal{F}(\mF_A,\mF_B) = \{(\mA,\mB) : [\mA] = \mF_A, [\mB] = \mF_B \}$. The operator $[\mA]$ represents a binarization of the matrix $\mA$. For this class, structural controllability is described as in Definition \ref{def:StrucCont} \cite{ching-tailinStructuralControllability1974,guoActuatorPlacementStructural2021a}. This definition implies that while full controllability may not be achieved under certain system scenarios, by manipulating the injection inputs and/or hydraulics, such as flow rates, full controllability can be achieved or at least approached and maximized.

\begin{mydef}~\label{def:StrucCont}
	A class of the system $\mathcal{F}(\mF_A,\mF_B)$ is structurally controllable if there exists at least one system which is controllable.
\end{mydef}

A class $\mathcal{F}$ of the system \eqref{eq:WQSS} is \textit{structurally controllable} is and only if the following conditions hold:
\begin{enumerate}
	\item $\mathcal{F}$ is \textit{input connected}: there exists at least one path to the state node $x_i$ which connects an input node $u_j, \; , \; \forall j= 1, \cdots, n_u$ with the state node $x_i, \; \forall i= 1, \cdots, n_x$.
	\item $\mathrm{s\text{-}rank}(\mF_A \; \mF_B) = n_x$: $\mathrm{s\text{-}rank}$ is the structural rank of a matrix and it is defined as the number of the one-elements in matrices which may be chosen in a way that they appear in different rows and columns. 
\end{enumerate}


In our study, we check if the system of $(\mA, \mB_\mathcal{S})$ is structurally controllable or not by using the SALS toolbox on MATLAB \cite{geiselSALSToolboxVersion2021}. We denote this check function as $\text{\texttt{sc}}(\mA, \mB_\mathcal{S})$ (refer to Step 6 in Algorithm \ref{alg:forward_greedy}).

Our strategy defines and assigns weights differently from traditional graph-theoretic approaches, which often assign weights to edges or, in some cases, to individual nodes. Instead, we assign these weights specifically to sets of candidate nodes, not individual locations or edges, to reflect the collective performance and strategic importance of these sets. These node-set weights are determined based on the following: \textit{(i)} how frequently the set is chosen and whether it achieves structural controllability, \textit{(ii)} assigning higher weights to sets that include locations frequently appearing in other sets during different time periods or scenarios, and \textit{(iii)} accounting for temporal patterns by giving higher weights to sets that are chosen during critical hours (e.g., peak water demand hours) or dominant hydraulic scenarios. Having said that, the weight of every set of booster station locations is expressed as in Eq. \eqref{eq:SetWeight}. Then, the set with the highest weight is selected.
\begin{equation}\label{eq:SetWeight}
	\begin{split}
		w_{\mathcal{S}_i} & = \mu_1 \frac{\sum_{j=1}^{N_\mathrm{HP}} |\{\mathcal{S}_i \subset \boldsymbol{\mathcal{S}}_j \}|}{N_\mathrm{HP} \times T_\mathrm{s}} \\
		& + \mu_2 \frac{\sum_{j=1}^{N_\mathrm{HP}} |\{\mathcal{S}_i \subset \boldsymbol{\mathcal{S}}_j \} \; | \; \vz(i) = 1 |}{\sum_{j=1}^{N_\mathrm{HP}} |\{\mathcal{S}_i \subset \boldsymbol{\mathcal{S}}_j \}|} \\
		& + \mu_3 \frac{\sum_{j=1}^{N_\mathrm{HP}} \sum_{k \in \mathcal{S}_i} |\{k \in \boldsymbol{\mathcal{S}}_j \}|}{n_s \times N_\mathrm{HP} \times T_\mathrm{s}} + \mu_4,
	\end{split}
\end{equation}
where $w_{\mathcal{S}_i}$ is the weight for each set $\mathcal{S}_i$ obtained by solving the CBSP problem for $N_\mathrm{HP}$ scenarios of hydraulic profiles, each with a total simulation period of $T_\mathrm{s}$. The term $|\{\mathcal{S}_i \subset \boldsymbol{\mathcal{S}}_j \}|$ counts how many times 
$\{\mathcal{S}_i$ appears as a subset in the collection $\boldsymbol{\mathcal{S}}_j$ for scenario $j$. The $|\cdot|$ symbol refers to the cardinality (or size) of the set. The second term $|\{\mathcal{S}_i \subset \boldsymbol{\mathcal{S}}_j \} \; | \; \vz(i) = 1 |$ counts how many times the set $\mathcal{S}_i$ is chosen and structural controllability is achieved for scenario $j$. Lastly, the term $\sum_{k \in \mathcal{S}_i} |\{k \in \boldsymbol{\mathcal{S}}_j \}|$ accounts for how often the individual elements (locations) within the set $\mathcal{S}_i$ appear across the scenarios and time steps. Moreover, $\mu_1, \mu_2,$ and $\mu_3$ are weighting coefficients used to prioritize one weighting term over the others, while $\mu_4$ is a weighting coefficient applied to the sets chosen during critical periods of the simulation.

Note that these weights provide us with the geographic locations of the \textit{fixed} booster stations, based on scenarios from the early stages of WDN operation. However, these scenarios may differ from the actual dynamics and conditions of the system. This is why we consider multiple possible common and realistic hydraulic profiles, and also account for temporary locations with flexibility to accommodate chlorine injections using mobile or fixed stations after observing network trends.

\begin{table}[h!]
	\centering
	\begin{threeparttable}
		\caption{Test networks components.}~\label{tab:NetworksComponents}
		\setlength\tabcolsep{.75\tabcolsep}%
		\begin{tabular}{c|c|c|c|c|c|c}
			\hline
			Network & Junctions & Reservoirs & Tanks & Pipes & Pumps & Valves \\
			\hline
			Three-node & 1 & 1 & 1 & 1 & 1 & 0 \\
			Net1 & 9 & 1 & 1 & 12 & 1 & 0 \\
			FOS & 36 & 1 & 0 & 58 & 0 & 0 \\
			Net3 & 90 & 2 & 3 & 114 & 2 & 0 \\
			C-Town & 388 & 1 & 7 & 429 & 11 & 4 \\
			\hline
			\hline
		\end{tabular}
	\end{threeparttable}
\end{table}

\section{Case Studies}~\label{sec:CaseStudies}
In this section, we validate the proposed booster station placement approach on various networks with different layouts, scales, characteristics, and typologies. These networks include the three-node network, Net1, FOS, Net3, and C-Town networks \cite{marchiBattleWaterNetworks2014,rossmanEPANET22User2020}. Fig. \ref{fig:CaseStudy} and Tab. \ref{tab:NetworksComponents} demonstrate the layout and number of components for each network. In the following sections, we showcase our approach to solve the CBSP problem applicability, validity, and scalability. In addition, we provide answers to the following specific questions:

\noindent \mybox[fill=copper!60]{\textbf{Q1}}$\textcolor{copper!60}{\blacktriangleright}$ How different are the placements when using the two different controllability metrics ($\log \det$ and $\trace$)?

\noindent \mybox[fill=copper!40]{\textbf{Q2}}$\textcolor{copper!40}{\blacktriangleright}$
How does our approach determine the final station configuration when structural controllability is rarely achieved?

\noindent \mybox[fill=copper!20]{\textbf{Q3}}$\textcolor{copper!20}{\blacktriangleright}$
How scalable are the proposed placement strategies?

\noindent \mybox[fill=copper!10]{\textbf{Q4}}$\textcolor{copper!10}{\blacktriangleright}$
How can the proposed placement be tailored for placement of mobile booster stations?

\subsection{Fixed Booster Stations Placement}
\begin{figure}[t]
	\centering
	\includegraphics[width=0.5\textwidth]{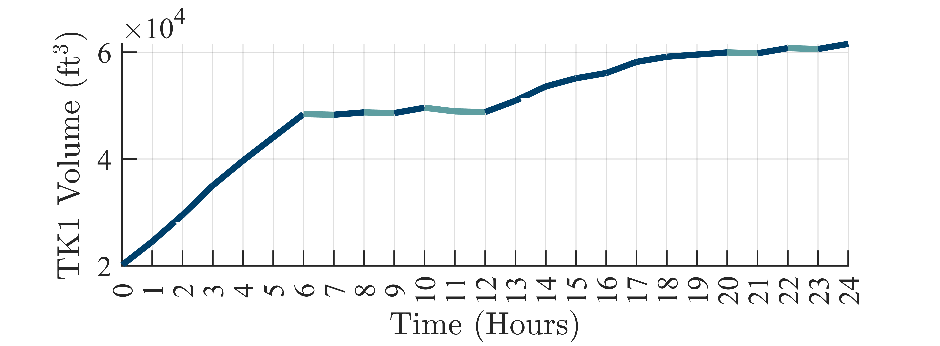}
	\vspace{-0.4cm}
	\caption{Tank TK1 volume of the Three-node network plotted with a darker color for windows when the tank is filling and lighter color for windows when it is emptying, over a total duration of 24 hours.}~\label{fig:3NTK1Vol}
 \vspace{-0.3cm}
\end{figure}

First, we solve the two formulated CBSP problems for the simple three-node network to analyze and compare the results. We solve for the hydraulic profile that results in the Tank TK1 volume illustrated in Fig. \ref{fig:3NTK1Vol}. Placements are determined for each hydraulic time-step of 1 hour, while the WQ time-step is set to 10 seconds over a total simulation period of $T_\mathrm{s}=24$ hours. For this network, there are three possible station locations (R1, J1, and TK1); thus, the placement problems are solved for $n_s=1$ and $n_s=2$. 

\begin{figure}[h!]
	\centering	\subfloat[\label{fig:3NBSt_Trace}]{\includegraphics[keepaspectratio=true,width=0.38\textwidth]{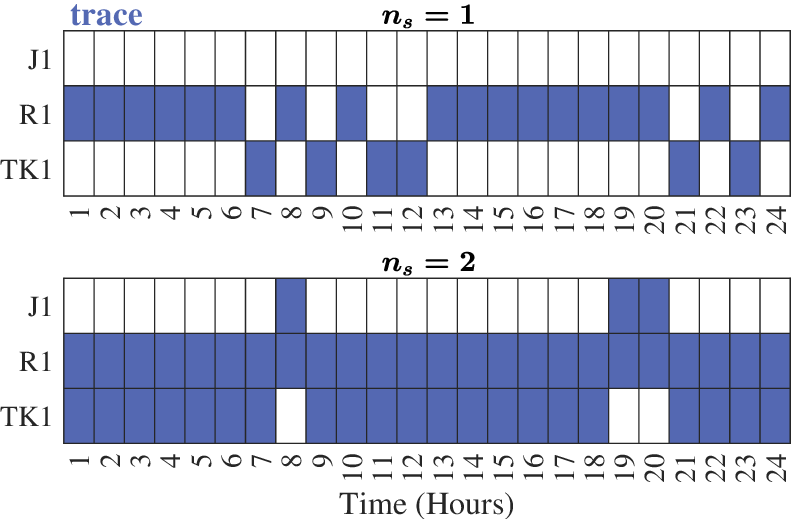}}{}\vspace{-0.25cm}
	\subfloat[\label{fig:3NBSt_LogDet}]{\includegraphics[keepaspectratio=true,width=0.38\textwidth]{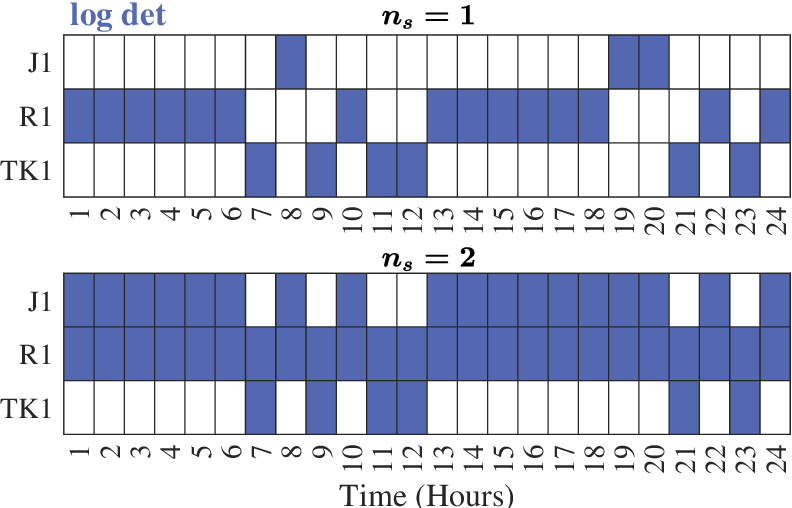}}{}\vspace{-0.15cm}
	\caption{Booster stations placement results for the Three-node network by solving the (a) $\trace$- and (b) $\log \det$-based CBSP problems for a simulation period of 24 hours with $\Delta t_\mathrm{WQ}=10$ sec and $\Delta t_\mathrm{H} = 1$ hour.}~\label{fig:3NBSt}
\end{figure}

The final CBSP results are presented in Fig. \ref{fig:3NBSt_Trace} for the $\trace$-based problem and in Fig. \ref{fig:3NBSt_LogDet} for the $\log \det$-based problem. When $n_s=1$, we observe that during periods where the tank is providing the system with water to satisfy the demand at J1, the booster station is located at Tank TK1 for both problems. Conversely, during tank filling periods, the $\trace$- and $\log \det$-based problems select R1 for the majority of the simulation period, with the $\log \det$ problem occasionally selecting J1, particularly during intervals of lower velocity in P1. When  $n_s=2$, the $\trace$-based problem favors TK1 over J1 unlike the $\log \det$-based problem, with R1 being a constant selection as the other placement location. This preference reflects the $\trace$ metric’s aim to maximize average control energy, while the $\log \det$ metric seeks broader state coverage, which is highlighted during lower velocity intervals. 

\begin{figure}[h!]
	\centering	\subfloat[\label{fig:Net1BSt_BD}]{\includegraphics[keepaspectratio=true,width=0.4\textwidth]{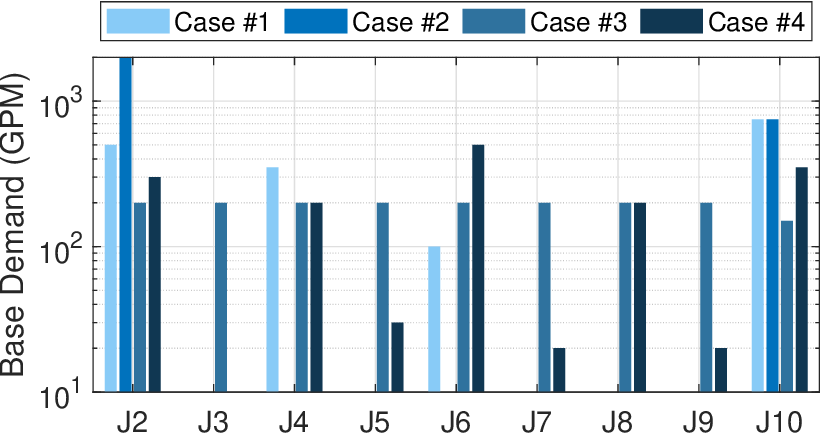}}{}\vspace{-0.3cm}
	\subfloat[\label{fig:Net1BSt_Pat}]{\includegraphics[keepaspectratio=true,width=0.4\textwidth]{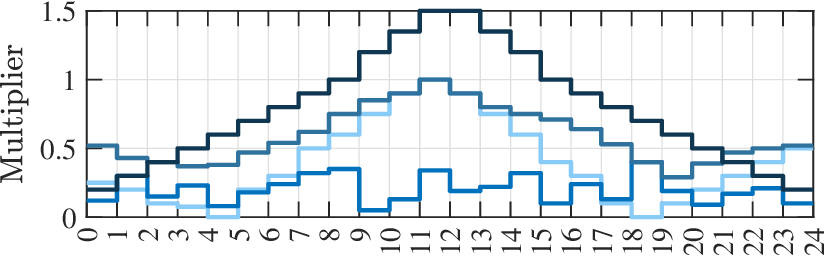}}{}\vspace{-0.45cm}
	\subfloat[\label{fig:Net1BSt_TK}]{\includegraphics[keepaspectratio=true,width=0.4\textwidth]{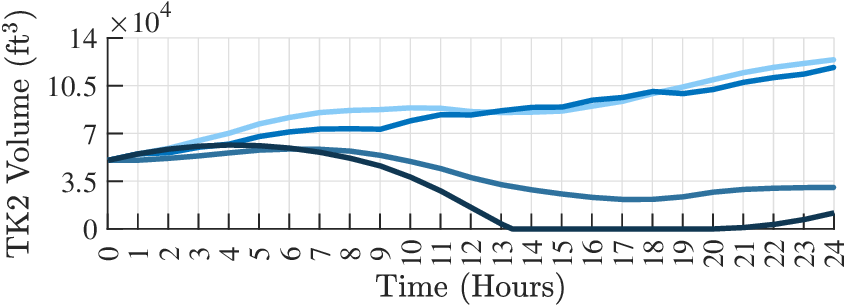}}{}\vspace{-0.2cm}
	\caption{Four case scenarios based on different (a) base demands, (b) demand patterns for nodes, and (c) the corresponding TK2 volume in Net1.\vspace{-2cm}}~\label{fig:Net1BSt_Cases}
\end{figure}

\begin{figure*}[h!]
	\centering
	\vspace{-0.7cm}	{\captionsetup[subfloat]{labelformat=empty}\subfloat[\label{fig:Net1_LogDet_k1}]{\includegraphics[keepaspectratio=true,width=0.83\textwidth]{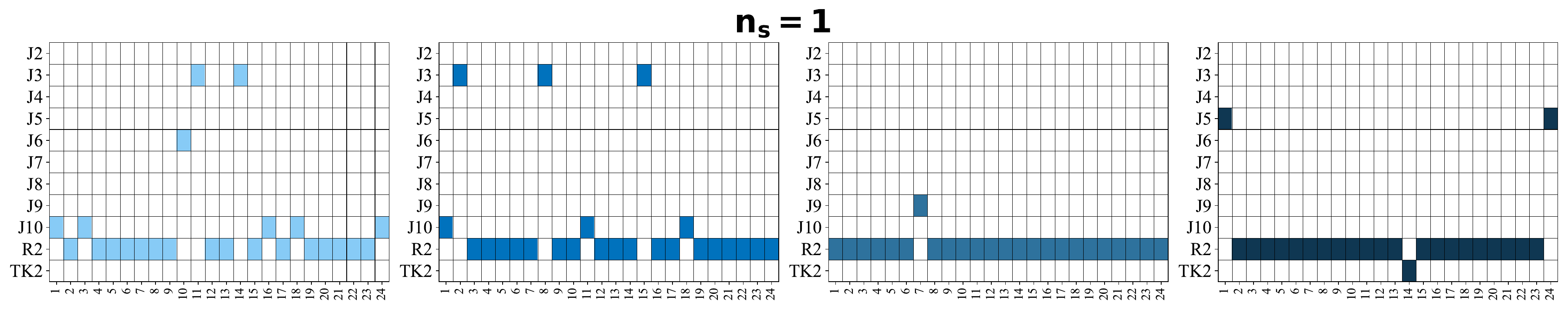}}{}\vspace{-1cm}}
	{\captionsetup[subfloat]{labelformat=empty}\subfloat[\label{fig:Net1_LogDet_k3}]{\includegraphics[keepaspectratio=true,width=0.83\textwidth]{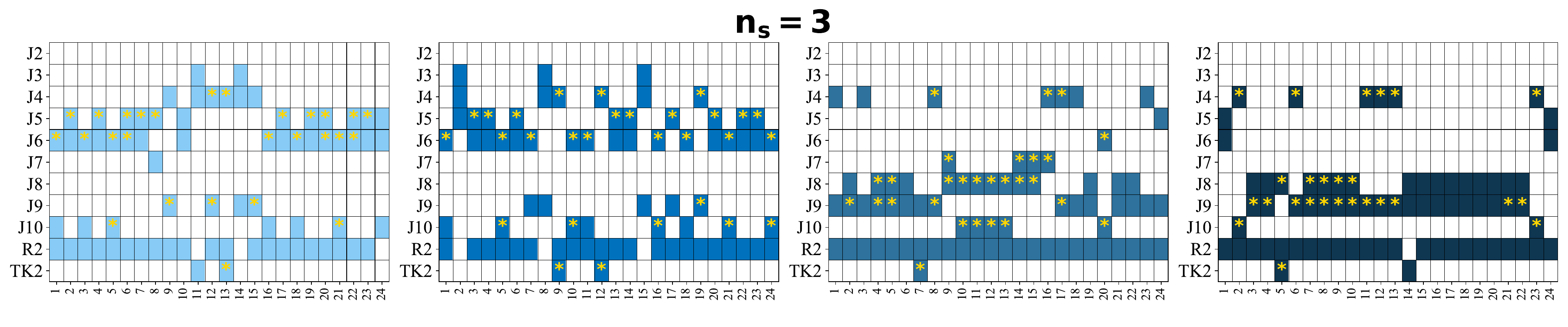}}{}\vspace{-1cm}}
	{\captionsetup[subfloat]{labelformat=empty}\subfloat[\label{fig:Net1_LogDet_k5}]{\includegraphics[keepaspectratio=true,width=0.83\textwidth]{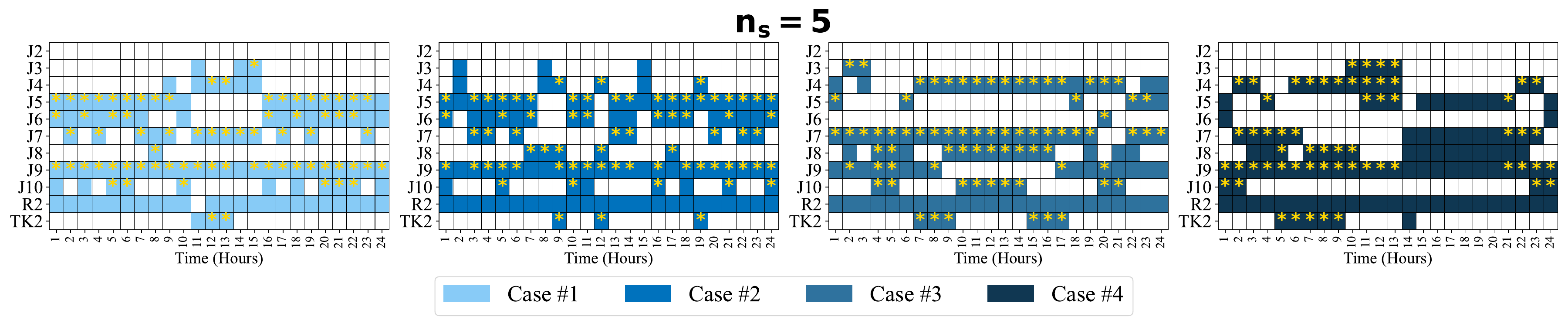}}{}\vspace{-0.6cm}}
	\caption{Booster stations placement results for Net1 network by solving the $\log \det$-based CBSP problem for a 24 hours with $\Delta t_\mathrm{WQ}=10$ sec and $\Delta t_\mathrm{H} = 1$ hour, for each of the four case scenarios. Stars indicate that the system is structurally controllable after placing booster stations at those specific nodes.\vspace{-0.8cm}}~\label{fig:Net1BStLogDet}
\end{figure*}

\begin{figure*}[h!]
	\centering
	\vspace{-0.4cm}
	{\captionsetup[subfloat]{labelformat=empty}\subfloat[\label{fig:Net1_Trace_k1}]{\includegraphics[keepaspectratio=true,width=0.83\textwidth]{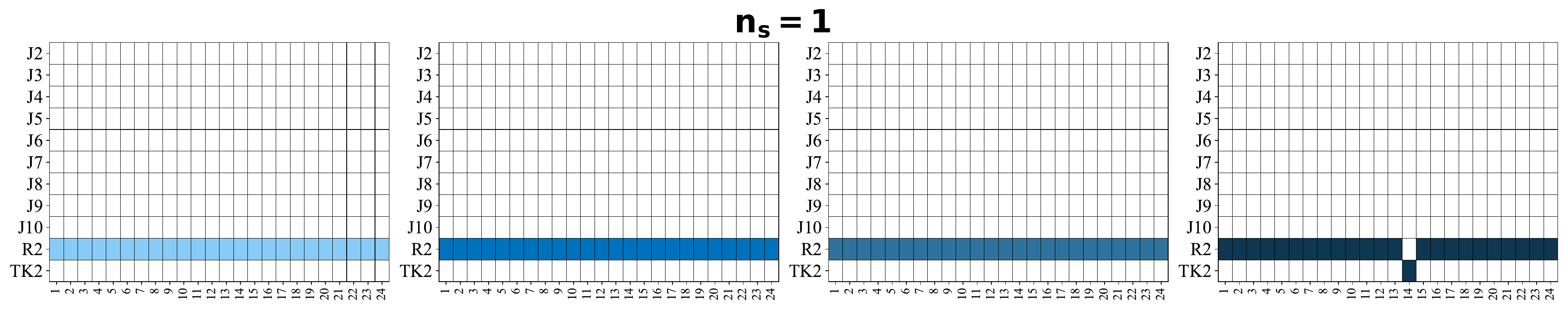}}{}\vspace{-1cm}}
	{\captionsetup[subfloat]{labelformat=empty}\subfloat[\label{fig:Net1_Trace_k3}]{\includegraphics[keepaspectratio=true,width=0.83\textwidth]{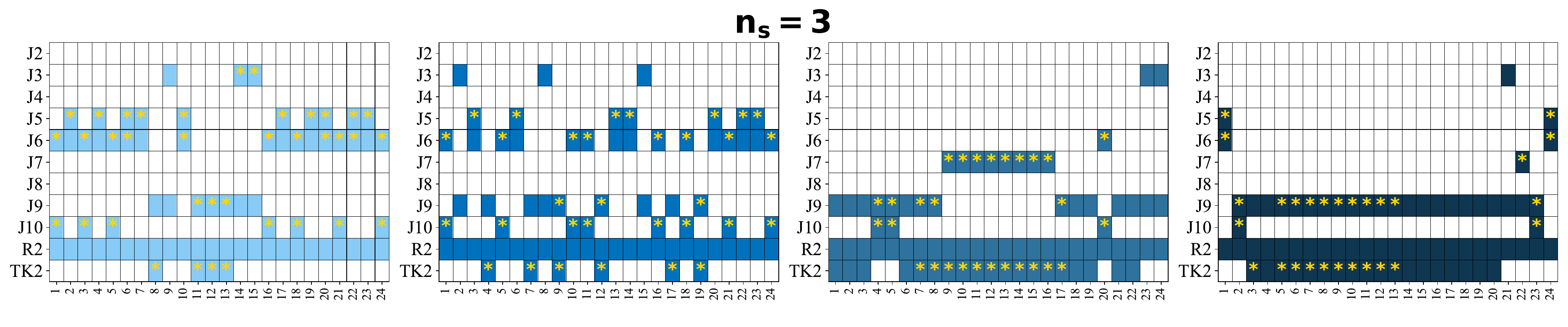}}{}\vspace{-1cm}}
	{\captionsetup[subfloat]{labelformat=empty}\subfloat[\label{fig:Net1_Trace_k5}]{\includegraphics[keepaspectratio=true,width=0.83\textwidth]{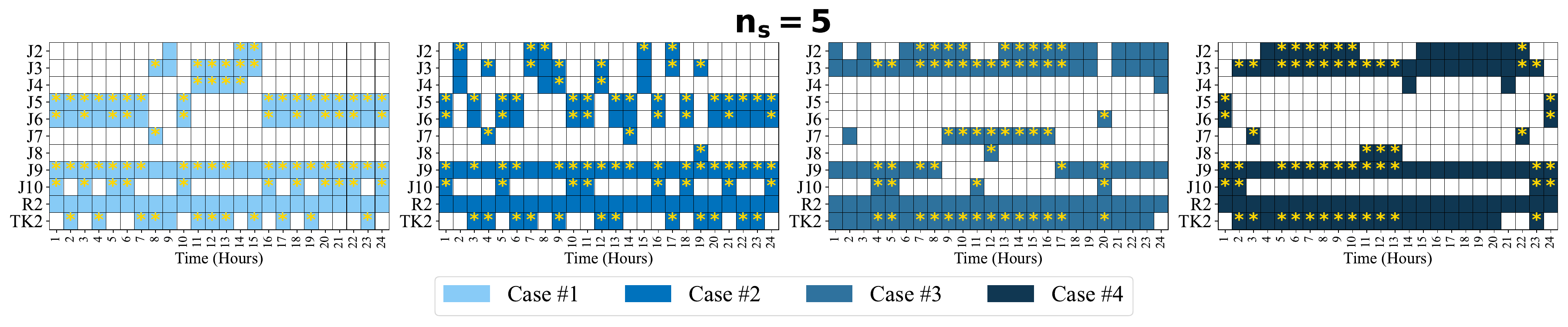}}{}\vspace{-0.5cm}}
	\caption{Booster stations placement results for Net1 network by solving the $\trace$-based CBSP problem for a 24 hours with $\Delta t_\mathrm{WQ}=10$ sec and $\Delta t_\mathrm{H} = 1$ hour, for each of the four case scenarios. Stars indicate that the system is structurally controllable after placing booster stations at those specific nodes.}~\label{fig:Net1BStTrace}
\end{figure*}

\begin{figure*}[h!]
	\centering
	\vspace{-0.5cm}
	\includegraphics[width=0.83\textwidth]{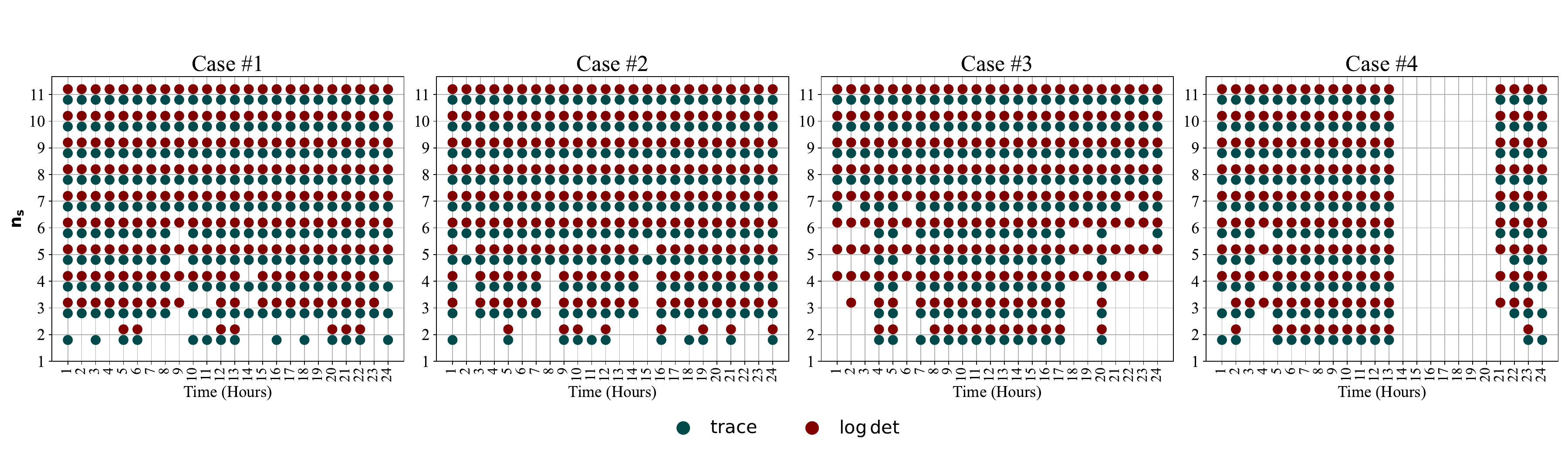}
	\vspace{-0.5cm}
	\caption{Structural controllability check for the four case scenarios of Net1, each spanning 24 hours, obtained by solving both $\trace$- and $\log \det$-based CBSP problems.}~\label{fig:Net1_StCh}
\end{figure*}

Furthermore, we adopt four varying demand patterns for Net1 with different base demands as illustrated in Fig. \ref{fig:Net1BSt_BD} and Fig. \ref{fig:Net1BSt_Pat}. As shown in Fig. \ref{fig:Net1BSt_TK}, the resulting changes in TK2's volume for these four case scenarios differ during the filling and emptying windows of the system. Under these scenarios, we determine the booster station locations for $n_s=1$, $n_s=3$, and $n_s=5$ by solving the $\trace$- and $\log \det$-based problems, with the results displayed in Fig. \ref{fig:Net1BStLogDet} and Fig. \ref{fig:Net1BStTrace}, respectively. In these figures, we annotate the structural controllability condition satisfaction 
by adding this specific node to the booster station set by a star. In addition, we present the results of the final structural controllability check for the system corresponding to the number of stations $n_s$ for both problems and all four case scenarios in Fig. \ref{fig:Net1_StCh}. All these results for all case scenarios are obtained by considering a WQ time-step of 10 secs and a hydraulic time-step of 1 hour.

These results demonstrate the submodularity property of the solutions for both problems, as the location set for $n_s=1$ is a subset of the location set for $n_s=3$, which in turn is a subset of the location set for $n_s=5$. Additionally, we observe that the two CBSP problems formulated in this study give different results, yet, achieve structural controllability commonly in most cases, indicating full control coverage. This is due to the looped topology of this network. This is attributed to the looped topology of the network. However, structural controllability is strictly unachievable during certain time windows, even when a booster station is hypothetically placed at every node in the network—as shown in Fig. \ref{fig:Net1_StCh}. This is due to the low velocities in some pipes, which make it difficult to cover the entire pipe length within a single hydraulic time-step, thereby the system does not check as controllable. One more thing to observe, the results in Fig. \ref{fig:Net1_StCh} provide insights into how many stations to achieve full coverage over the system under specific scenarios.

\begin{table}[h!]
	\centering
	\vspace{-0.25cm}
	\begin{threeparttable}
		\caption{{Booster stations placement results for Net1 under three weighting scenarios and by solving the two CBSP problems.}}~\label{tab:Net1DiffWeights}
		\setlength\tabcolsep{1.2\tabcolsep}%
		\begin{tabular}{c|c|c}
			\hline
			Weighting Scenario & $\log \det$-based & $\trace$-based \\
			\hline
			WS\#1 & R2 \& J6 \& J10 & R2 \& J5 \& J6  \\
			WS\#2   & R2 \& J8 \& J9 & R2 \& J9 \& TK2  \\
			WS\#3 & R2 \& J6 \& J9 & R2 \& J9 \& TK2 \\
			\hline
			\hline
		\end{tabular}
	\end{threeparttable}
\end{table}

Next, we apply the weight strategy proposed in Section \ref{sec:BStVsHydraulic} to the results obtained for Net1 by solving the two CBSP problems. First, no higher weights are given to any hydraulic profile or simulation window and all the other weighting coefficients in \eqref{eq:SetWeight} are taken equal 1---this weighting scenario is referred to as WS\#1. In the second scenario, the weighting coefficients are modified to exclude structural controllability from the equation ($\mu_2=0$), resulting in weighting scenario WS\#2. In the final scenario, WS\#3, we select the locations that appear most frequently. Results from these three scenarios for the two CBSP problems are listed in Tab. \ref{tab:Net1DiffWeights}. We observe that the results vary depending on which metric the problem is based on and the weighting strategy applied. This conclusion produces the following answer to Q1.

\noindent \mybox[fill=copper!60]{\textbf{A1}}$\textcolor{copper!60}{\blacktriangleright}$
The $\log \det$-based CBSP problem produces different results than the $\trace$-based problem. This is due to the distinct interpretation of these two metrics and how they quantify WQ controllability, as explained in Section \ref{sec:WQControl}. The choice between these metrics depends on the specific operational needs of the water network. For WDN operators, the $\trace$ metric is suitable when the goal is to minimize the average energy required across the entire system, making it ideal for networks with more uniform flow patterns. On the other hand, the $\log \det$ metric should be chosen when the focus is on maximizing the reachable state space with minimal energy, which is beneficial for networks with high variability in demand and flow, ensuring that chlorine injections can cover a larger portion of the system with limited energy input.


Building on the previous answer, we evaluate the placements obtained from both the $\log \det$-based and $\trace$-based CBSP problems by applying the model predictive control (MPC) approach proposed in \cite{wangHowEffectiveModel2021a} to determine the chlorine injections needed to maintain residuals between 0.2 and 4 mg/L, as required by EPA regulations \cite{acrylamideNationalPrimaryDrinking2009}. Results show that both placement configurations perform well across the different hydraulic profiles, with trade-offs between the windows where full controllability is achieved and those where it is not, resulting in total chlorine injections that are close in amount.

   \begin{figure}[h!]
	\centering	\subfloat[\label{fig:Net1_GvsR_Trace_3}]{\includegraphics[keepaspectratio=true,width=0.41\textwidth]{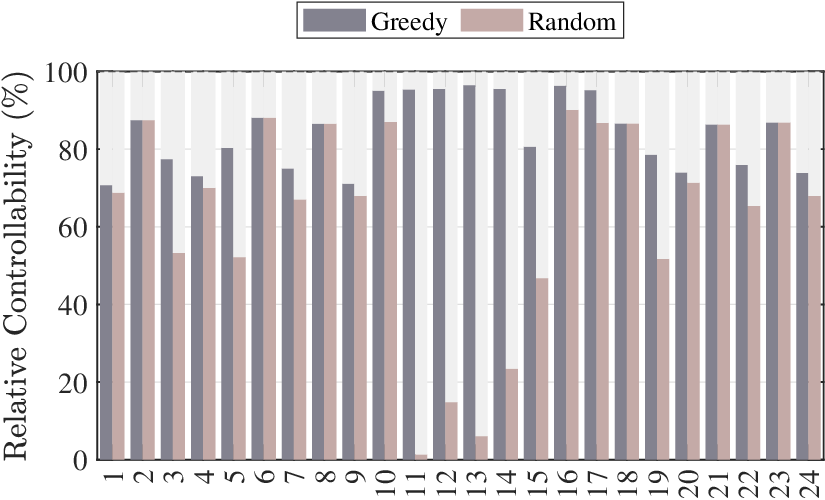}}{}\vspace{-0.25cm}
	\subfloat[\label{fig:Net1_GvsR_Trace_5}]{\includegraphics[keepaspectratio=true,width=0.41\textwidth]{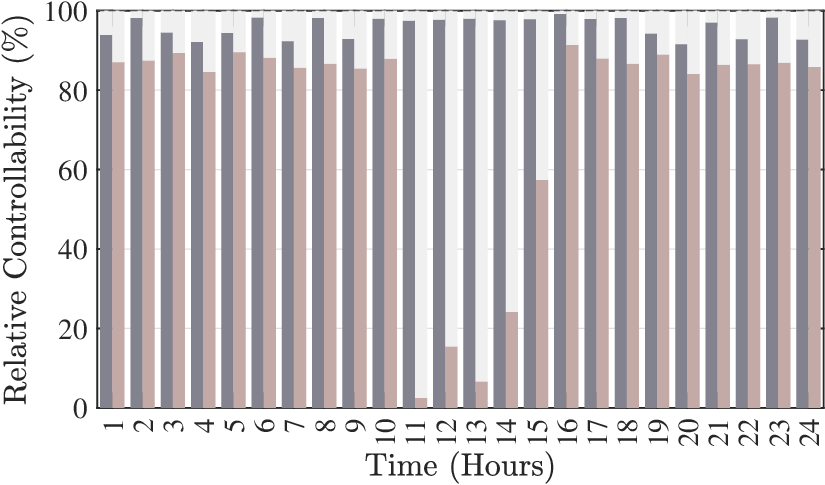}}{}\vspace{-0.1cm}
	\caption{Comparison of relative controllability achieved by optimized (Greedy) versus random booster station placements for the Net1 network under Case Scenario \#3, using the $\mathrm{trace}$-based controllability metric. Results are shown as percentages relative to full (uniform) actuation for (a) $n_s = 3$ and (b) $n_s = 5$ booster stations.}~\label{fig:Net1_GvsR_Trace}
\end{figure}

    \begin{figure}[h!]
	\centering	\subfloat[\label{fig:Net1_GvsR_LogDet_3}]{\includegraphics[keepaspectratio=true,width=0.41\textwidth]{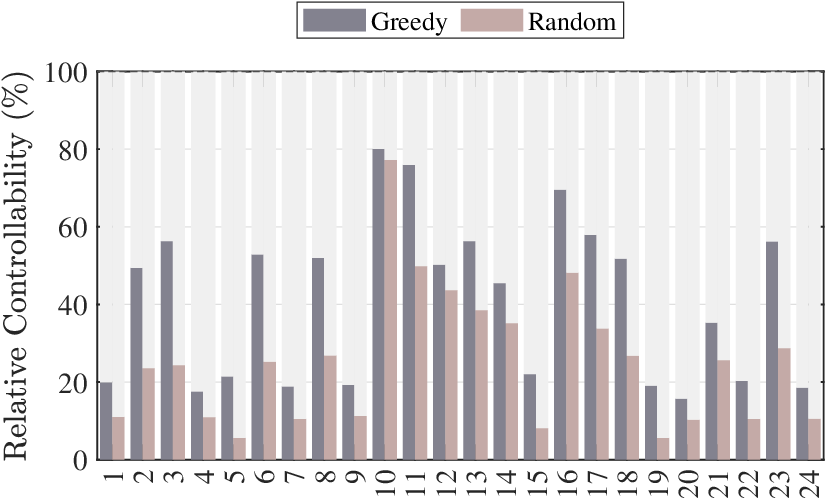}}{}\vspace{-0.25cm}
	\subfloat[\label{fig:Net1_GvsR_LogDet_5}]{\includegraphics[keepaspectratio=true,width=0.41\textwidth]{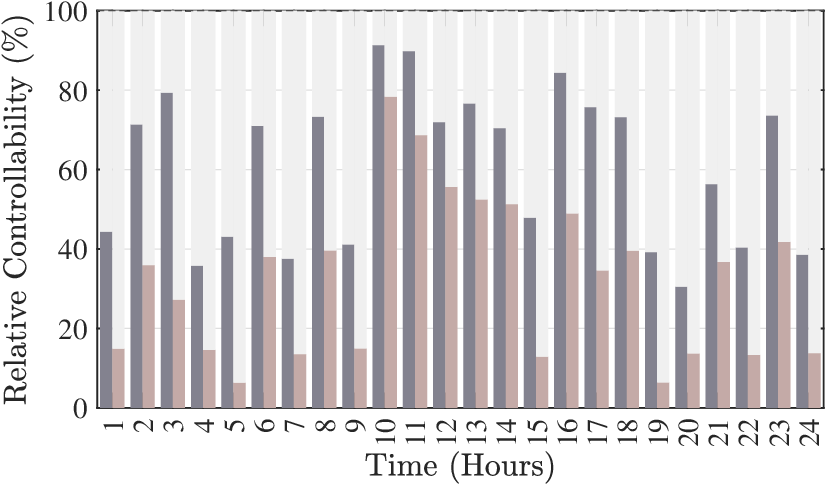}}{}\vspace{-0.1cm}
	\caption{Comparison of relative controllability achieved by optimized (Greedy) versus random booster station placements for the Net1 network under Case Scenario \#3, using the $\log \det$-based controllability metric. Results are shown as percentages relative to full (uniform) actuation for (a) $n_s = 3$ and (b) $n_s = 5$ booster stations.}~\label{fig:Net1_GvsR_LogDet}
\end{figure}

To evaluate the performance of the proposed placement method, we compare its results on Net1 with two alternative booster station placement strategies: a random strategy and a uniform strategy. In the random strategy, the fixed number of booster stations $n_s$ are placed at randomly selected candidate nodes. This process is repeated with different random seeds to obtain a representative distribution of outcomes. In the uniform strategy, all candidate nodes are actuated, representing an idealized case with full actuation and thus an upper bound on the controllability metric. For comparison, we introduce a normalized metric referred to as \textit{relative controllability (\%)}, which expresses the controllability achieved by each method as a percentage of the controllability achieved by the uniform strategy. That is, a value of 100\% corresponds to the maximum controllability obtainable when all candidate nodes are actuated, and values below this indicate the performance of the strategy relative to this upper bound.

Fig.~\ref{fig:Net1_GvsR_Trace} and Fig.~\ref{fig:Net1_GvsR_LogDet} illustrate this comparison for the hydraulic Case Scenario \#3 of Net1 for both $\trace$- and $\log \det$-based CBSP formulations, respectively, for $n_s=3$ and $n_s=5$ booster stations. The results shown are based on one realization of the random placement strategy for visualization purposes; however, we have repeated the random placement procedure 25 times using different random seeds and have consistently observed that the greedy algorithm outperformed the random strategy in every instance. This demonstrates the effectiveness of the greedy approach and its consistent ability to deliver better system controllability with a limited number of stations.

As observed, for the $\trace$-based placement, the relative controllability achieved by the greedy approach is notably close to the uniform allocation even with only $n_s = 3$ stations, and the performance improves further with $n_s = 5$. In contrast, the $\log \det$-based placement shows lower controllability with $n_s = 3$, but experiences a improvement when the number of stations increases to $n_s = 5$. This difference is expected and reflects the operational meaning of these two metrics as highlighted and emphasized by the answer to the previously posed question. The $\trace$ metric quantifies the average input energy required to control the system and is more sensitive to the overall distribution of control effort. In contrast, the $\log \det$ metric measures the volume of the reachable state space and is more influenced by spatial variability in flow, demand, and overall network connectivity.

\begin{table}[h!]
	\centering
	\vspace{-0.1cm}
	\begin{threeparttable}
		\caption{{Booster station placement results for FOS network for two hydraulic scenarios (H\#1 \& H\#2) with $n_s=5$, $\Delta t_\mathrm{WQ} = 10$ sec and $\Delta t_\mathrm{H} = 1$ hr.}}~\label{tab:FOSBStP}
		\setlength\tabcolsep{0.65\tabcolsep}%
		\begin{tabular}{c|c|c}
			\hline
		Scenario & $\log \det$-based & $\trace$-based \\
			\hline
			H\#1 & R3 \& J13 \& J14 \& J16 \& J19 & R3 \& J15 \& J16 \& J22 \& J23 \\
			H\#2   & R3 \& J15 \& J22 \& J23 \& J24 & R3 \& J12 \& J23 \& J24 \& J25 \\
			\hline
			\hline
		\end{tabular}
	\end{threeparttable}
	\vspace{-0.15cm}
\end{table}

Furthermore, we apply the proposed CBSP approach to the FOS and Net3 networks. These two networks differ in scale and layout, with FOS having a looped configuration, while Net3 contains more dead-ends and a greater number of components. Results for FOS are listed in Tab. \ref{tab:FOSBStP} for two hydraulic scenarios (H\#1 \& H\#2). These two hydraulic scenarios result in different flow directions, and for a looped network, this results in different patterns and dynamics of which node affects which with no dominant one. However, we have tested the selected set of one scenario for the other one and the relative change in the system controllability is less than 20\%.

\begin{figure}[t!]
	\centering
	\includegraphics[width=0.48\textwidth]{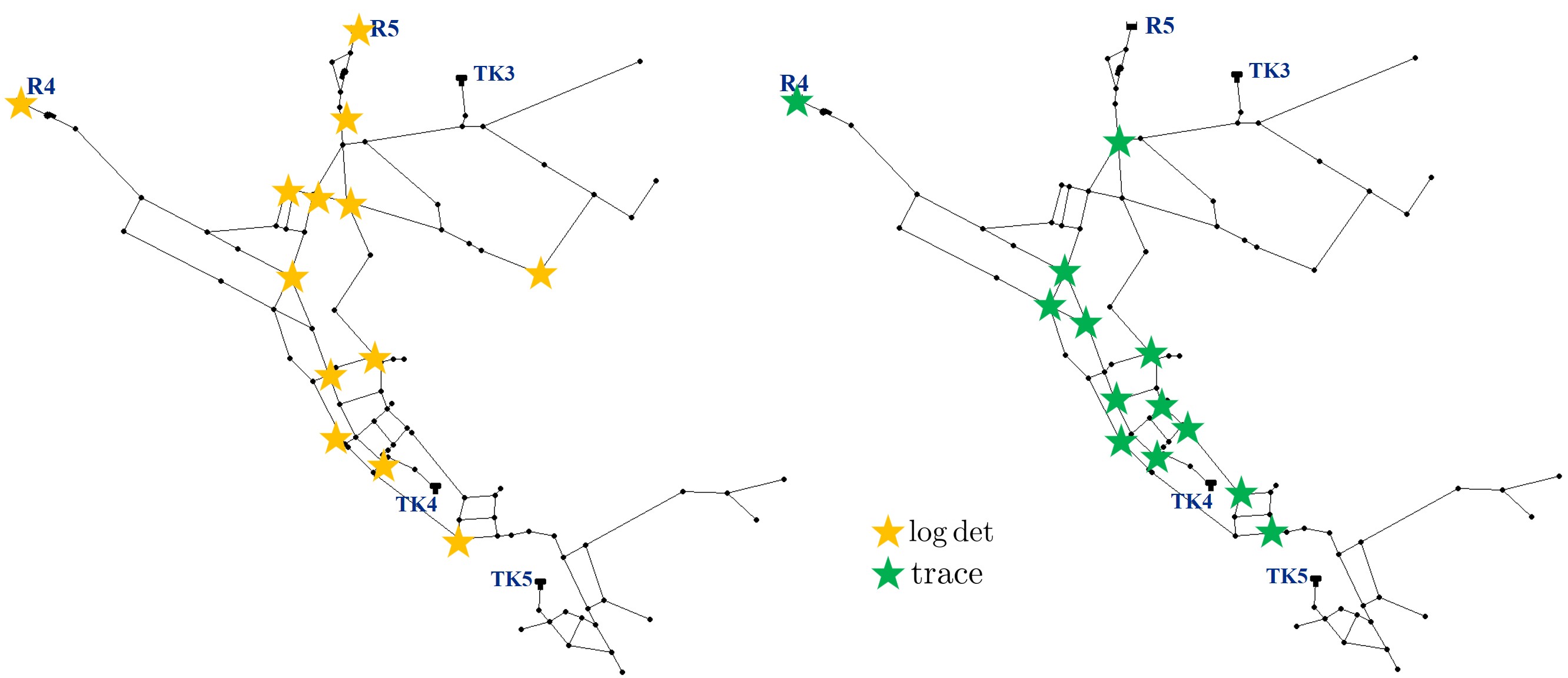}
	\caption{Booster station results for Net3 with $n_s=13$ by solving both problems and with $\Delta t_\mathrm{WQ} = 30$ sec and $\Delta t_\mathrm{H} = 1$ hr and by weighting the stations set by structural controllability.}~\label{fig:Net3BStPsc}
	\vspace{-0.1cm}
\end{figure}

\begin{figure}[t!]
	\centering
	\includegraphics[width=0.48\textwidth]{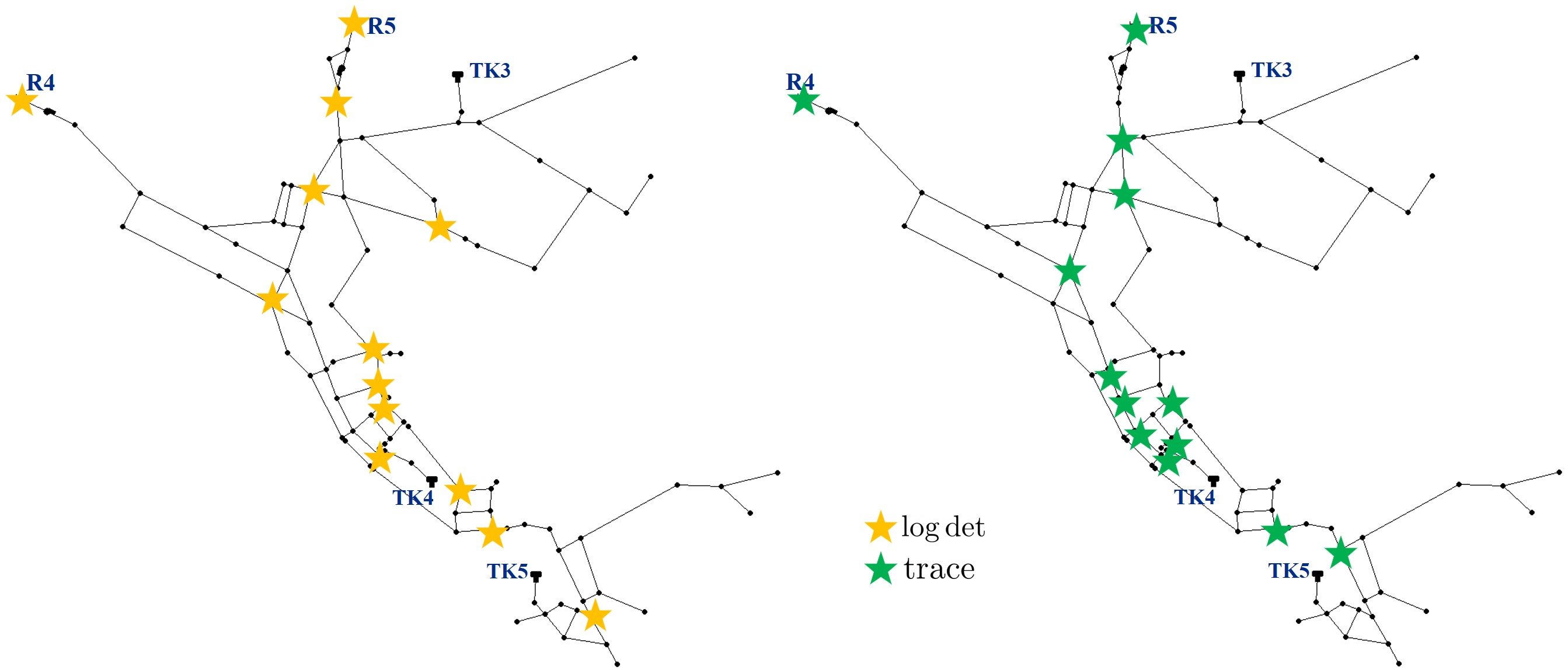}
	\caption{Booster station results for Net3 with $n_s=13$ by solving both problems and with $\Delta t_\mathrm{WQ} = 30$ sec and $\Delta t_\mathrm{H} = 1$ hr and by weighting the dimension of the structurally reachable subspace.}~\label{fig:Net3BStPdim}
	\vspace{-0.1cm}
\end{figure}

Although FOS is a looped network, structural controllability is not achieved for many time windows with limited numbers of booster stations: $n_s=5$. This issue is even more persistent in networks with several dead-ends, as is the case with Net3, with the booster station placements illustrated in Fig. \ref{fig:Net3BStPsc}. These placements rarely achieve full controllability due to changing flow directions and nodes with only a single path to them. Under such conditions, our weighting strategy selects the set most frequently chosen. To address this, we adaptively provide an alternative for assessing each set based on the \textit{dimension of the structurally reachable subspace}. In other words, we assess what percentage of the system’s states are structurally controllable by their connectivity to the booster stations in each set. This can be achieved by using the $\texttt{dimsrs}(\mA,\mB_\mathcal{S})$ command from the SALS toolbox, applied after determining the optimal set in the greedy algorithm (after Step 8 in Algorithm \ref{alg:forward_greedy}). After applying this approach, the final placements for Net3 are presented in Fig. \ref{fig:Net3BStPdim}. This answers the  second question Q2. 

\noindent \mybox[fill=copper!40]{\textbf{A2}}$\textcolor{copper!40}{\blacktriangleright}$ In cases where structural controllability is never or rarely achieved with the obtained booster station placements, our approach can be adaptively expanded to evaluate the sets based on the size of the structurally reachable subspace.

\begin{figure}[h!]
	\centering
	\includegraphics[width=0.43\textwidth]{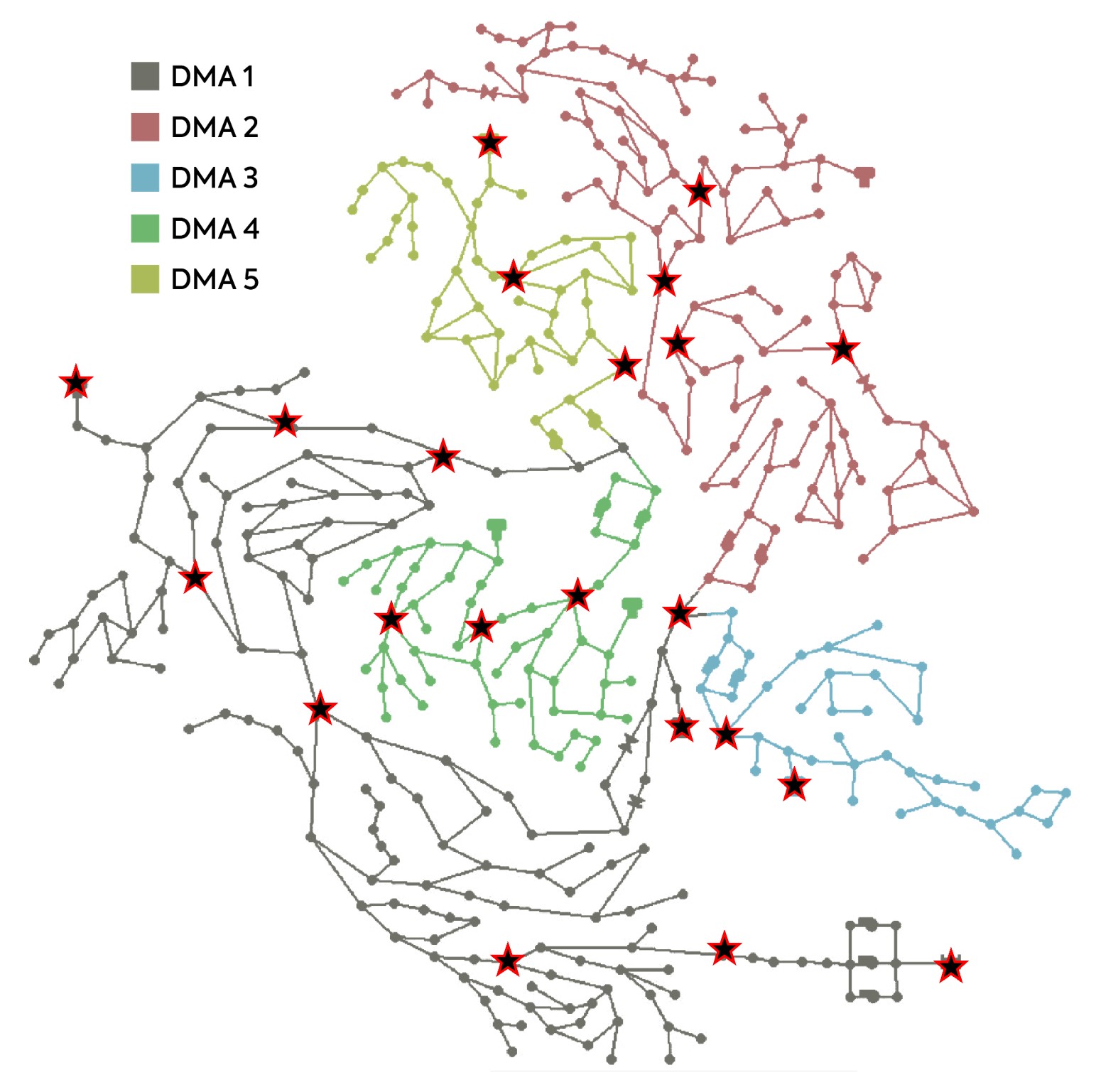}
	\caption{Booster station placements, indicated by stars, for C-Town network and the $\trace$-based CBSP problem with $\Delta t_\mathrm{WQ} = 30$ sec and $\Delta t_\mathrm{H} = 1$ hr.}~\label{fig:CTownBStP}
		\vspace{-0.3cm}
\end{figure}

The C-Town network is considered to be divided into five district metered areas (DMAs), as provided in \cite{marchiBattleWaterNetworks2014}. Each DMA has its own pumping system, with DMA1 serving as the main connector to the other four areas. Solving the CBSP problem for the C-Town network is challenging due to the high number of states, particularly after discretizing each pipe, resulting in thousands of states. Applying the greedy algorithm to select nodes from such a large set at every time step, while measuring the controllability metric for a system of this scale, is computationally demanding, if not infeasible. To address this issue, we divide the network into five subspaces and solve the CBSP problem separately for each, while accounting for how these subspaces interact. This is achieved by building the state matrix $\mA$ for each section of the network and incorporating connections to other sections that influence them into the $\mB$ matrix, with the associated flow rates under the assumption that water coming from these sections is already chlorinated. Therefore, we treat the C-Town network as five sub-networks, with DMA1 influencing the remaining four. The results of solving the $\trace$-based CBSP problem are shown in Fig. \ref{fig:CTownBStP}. It is important to note that, although booster stations in DMA1 affect the rest of the network, they are only evaluated for their coverage of DMA1, under the assumption that these areas are operated and managed independently. However, in networks without clear separations, these placements can be assessed across broader subspaces. This approach provides the answer to the third research question Q3. 

\noindent \mybox[fill=copper!20]{\textbf{A3}}$\textcolor{copper!20}{\blacktriangleright}$ The proposed CBSP approach can be scaled to large networks by dividing the system into smaller, manageable subspaces and solving the problem for each subspace independently. While network partitioning is outside the scope of this paper, readers are referred to \cite{khoa2020water,chanfreutSurveyClusteringMethods2021} for further discussion on this topic. This method considers the interactions between subspaces by adjusting the system matrices to reflect flow connections, ensuring that each section is treated with minimal computational complexity. By using this strategy, the approach remains computationally feasible while maintaining practical applicability in large-scale networks.

\subsection{Time-steps and Scales Sensitivity Analysis}
In this section, we test the sensitivity of the results to the chosen WQ and hydraulic time-steps. The choice of the WQ time-step influences the number of segments each pipe is divided into according to the upwind discretization scheme (refer to Section \ref{sec:WQModel} and Eq. \ref{eq:PipeWQ}). In addition, in our approach, we define the target controllable time, for which we solve the CBSP problem, as the hydraulic time-step.

First, we resolve the two CBSP problems for Net1 under the four case scenarios introduced in Fig. \ref{fig:Net1BSt_Cases}, with $\Delta t_\mathrm{WQ}=30$ sec instead of $10$ sec---this scenario is denoted as Case WQ30. Second, we change the hydraulic time-step $\Delta t_\mathrm{H}$ from 1 hour to 30 min, and refer to this scenario as Case H30. Last, Case WQH30 combines both changes, with $\Delta t_\mathrm{WQ}=30$ sec and $\Delta t_\mathrm{H}=30$ min. The final placements for these scenarios with $n_s=3$ and using the same weighting scenarios as the Base Case WS\#1 are listed in Tab. \ref{tab:Net1DiffT}. As observed, results differ with changes in time-steps and problem scale. However, the weights of the obtained sets remain close to those selected for the base case. In Case WQ30, the number of segments is reduced, lowering the system’s dimensionality and improving computational efficiency. In return, the average relative change in the controllability metric across the four hydraulic scenarios is approximately 13.2\%, which remains acceptable. On the other hand, although case H30 yields the same results as the base scenario, it might introduce an imbalance in the weighting system. This is because demand patterns change over a wider window than this hydraulic time-step and the booster station locations are determined before the chlorine injections have been fully distributed through the network, causing misalignment with the hydraulic updates. Additionally, reducing the hydraulic time-step increases computational demand, as the problem is solved twice as often compared to the base case. Therefore, the hydraulic time-step---and by extension, the window within which system controllability is assessed and locations are determined---should be chosen to reflect intervals where significant hydraulic changes, such as flow direction and magnitude, occur.

\begin{table}[t]
	\centering
	\vspace{-0.1cm}
	\begin{threeparttable}
		\caption{{Sensitivity results for booster station placements across four case scenarios: Base Case ($\Delta t_\mathrm{WQ}=10$ sec, $\Delta t_\mathrm{H}=1$ hour); Case WQ30 ($\Delta t_\mathrm{WQ}=30$ sec, $\Delta t_\mathrm{H}=1$ hour); Case H30 ($\Delta t_\mathrm{WQ}=10$ sec, $\Delta t_\mathrm{H}=30$ min); and Case WQH30 ($\Delta t_\mathrm{WQ}=30$ sec, $\Delta t_\mathrm{H}=30$ min).}}~\label{tab:Net1DiffT}
		\setlength\tabcolsep{1.2\tabcolsep}%
		\begin{tabular}{c|c|c}
			\hline
			Sensitivity Scenario & $\log \det$-based & $\trace$-based \\
			\hline
		Base Case WS\#1 & R2 \& J6 \& J10 & R2 \& J5 \& J6  \\
			Case WQ30   & R2 \& J10 \& TK2 & R2 \& J6 \& TK2 \\
			Case H30 & R2 \& J6 \& J10 & R2 \& J5 \& J6 \\
			Case WQH30 &  R2 \& J6 \& J10  & R2 \& J6 \& TK2 \\
			\hline
			\hline
		\end{tabular}
	\end{threeparttable}
	\vspace{-0.4cm}
\end{table}

\subsection{Backup Mobile Booster Stations Allocation}
In this section, we validate our answer to the fourth and final posed research question posed Q4.

\noindent \mybox[fill=copper!10]{\textbf{A4}}$\textcolor{copper!10}{\blacktriangleright}$ Yes, our approach can be used to determine backup locations for mobile chlorine injections under conditions of stations malfunctioning. By utilizing the greedy algorithm, the system's controllability with the existing booster stations is assessed, and then another node is selected that would contribute the most to improving this controllability. This node is chosen from a pool of candidate nodes, excluding those that already have stations or malfunctioning stations, while including nodes with accessibility for mobile chlorine injection.

For example, in the second hydraulic scenario of Net1 (Fig. \ref{fig:Net1BSt_Cases}), the selected set for $n_s=3$ by solving the $\trace$-based CBSP problem is $\mathcal{S}=\{$R2, J5, J6$\}$. If the booster station at J5 stops functioning at hour 12, we solve the problem again using Algorithm \ref{alg:forward_greedy} for the remaining 12 hours, and J9 is identified as the replacement location. Another approach is to solve the problem for only a few successive hours, allowing for temporary chlorine injection at the replacement location until the station is repaired.


\section{Conclusions, Limitations, and Future Work}~\label{sec:ConcLimFW}
This paper proposes a framework to address the CBSP problem in WDNs. The problem is formulated as a set function optimization problem, aiming to select booster station geographical locations that maximizes the WQ controllability. By framing the problem as a set function and incorporating submodular metrics of the WQ controllability Gramian, a forward greedy algorithm is employed to determine near-optimal solutions, with a guarantee on the deviation from optimality. We apply this approach to case studies involving networks of various scales and analyze the sensitivity of the approach to computational parameters and operational constraints. This work recognizes the practical challenges of applying the method to actual networks by offering operators a scalable version of the approach.    

This proposed approach can be expanded to include additional objectives beyond maximizing WQ controllability, such as those related to cost, social factors, and other considerations. Moreover, scenarios involving contamination intrusions can be incorporated, with specific attention given to zones that are particularly vulnerable to such events. Future work will focus on addressing these limitations to further develop the approach and to consider WQ multi-species dynamics (refer to \cite{elsherifControltheoreticModelingMultispecies2023}) in the process model, which will require utilizing methods and metrics different than the ones presenting in this study to assess and quantify controllability for a nonlinear system. This consideration is important as noted in~\cite{fisher2019limitations}, where the author points to limitations of methods that rely on single-species quality models with linear kinetics. In addition, chlorine reaction and decay rates, being uncertain and variable across seasons, sources, and zones, can be represented as distinct scenarios based on calibrated values obtained from field and historical data. The placement problem can then be solved independently under each scenario, yielding different candidate configurations. The weighting technique developed in this paper, grounded in structural controllability, is subsequently used to evaluate these configurations from a graphical and structural standpoint. This allows the identification of node sets whose placement remains effective across a range of plausible decay conditions, without relying on precise numerical agreement.

While this paper assumes a predefined number of booster stations, which is a realistic constraint in many practical settings, our framework remains flexible and can be extended to explore trade-offs between control performance and the cost of adding additional stations. This can be achieved by exploiting the submodular property of the controllability metrics through heuristic approaches that assess marginal gains in controllability as the number of stations varies. Alternatively, a more theoretical formulation has been proposed by our research group in \cite{kazmaMultilinearExtensionsSubmodular2025}, which expands on the approach adopted here by incorporating budget constraints directly into the placement optimization problem, an approach that builds on work presented in~\cite{calinescu2011maximizing,rezazadeh2023distributed}. Introducing such a constraint transforms the problem into one that requires different tools and optimization strategies, while still maintaining the control-theoretic foundation and offering submodular optimization performance guarantees. While this method has not yet been applied to the specific problem of chlorine booster station placement considered in this paper, it presents a promising direction for future work.


\balance
\bibliographystyle{IEEEtran}
\bibliography{WQ_BStPlacement}

\end{document}